\documentclass[aip,preprint,groupedaddress,showpacs]{revtex4}
\usepackage{graphicx}
\usepackage{mathrsfs}
\usepackage{color}
\usepackage{amsmath,amssymb}
\usepackage{dcolumn}
\newcommand{\br}{{\bf r}}
\newcommand{\bk}{{\bf k}}

\newcommand{\bq}{{\bf q}}
\newcommand{\bR}{{\bf R}}
\newcommand{\bE}{{\bf E}}

\newcommand{\bp}{{\bf p}}

\newcommand{\I}{{\rm i}}
\newcommand{\D}{\mathrm{d}}
\newcommand{\E}{\mathrm{e}}

\begin{document}
\title{Optical extinction, refractive index, and multiple scattering \\ for suspensions of interacting colloidal particles}
\author{Alberto Parola}
\affiliation{Department of Science and High Technology,
Universit\`{a} dell'Insubria, Via Valleggio 11, 22100 Como, Italy}
\author{Roberto Piazza}
\affiliation{Department of Chemistry (CMIC), Politecnico di
Milano, Via Ponzio 34/3, 20133 Milano, Italy}
\email{roberto.piazza@polimi.it}
\author{Vittorio Degiorgio}
\affiliation{Department of Industrial and Information Engineering,
Universit\`{a} di Pavia, Via Ferrata 5a, 27100 Pavia, Italy}
\date{\today}
\begin{abstract}
We provide a general microscopic theory of the scattering
cross-section and of the refractive index for a system of
interacting colloidal particles, exact at second order in the
molecular polarizabilities. In particular: a) we show that the
structural features of the suspension are encoded into the forward
scattered field by multiple scattering effects, whose contribution
is essential for the so-called ``optical theorem'' to hold in the
presence of interactions; b) we investigate the role of radiation
reaction on light extinction; c) we discuss our results in the
framework of effective medium theories, presenting a general
result for the effective refractive index valid, whatever the
structural properties of the suspension, in the limit of particles
much larger than the wavelength; d) by discussing
strongly-interacting suspensions, we unravel subtle anomalous
dispersion effects for the suspension refractive index.
\end{abstract}
\maketitle
\section{Statement of the problem}
\label{intro} Scattering methods have long been a basic tool for
the investigation of colloidal systems. The recent development of
optical correlation techniques\cite{Duri2009,Cerbino2008,
Buzzaccaro2013, Piazza2014} that, by successfully combining
scattering and real-space visualization, allow to probe the
microscopic Brownian dynamics still retaining the spatial
resolution proper of a microscope, calls however for a critical
reassessment of the relation between scattering and imaging. A
detailed analysis of the effects of the propagation through a
scattering medium  on the amplitude and phase of the transmitted
wavefront is also of primary importance for digital holographic
techniques\cite{Zhang1998}.

The effect on the transmitted wavefront of the transit though a
scattering medium can be expressed by stating that the forward
scattering pattern consists in a faithful reproduction of the
incident field that spatially superimposes with the transmitted
radiation, but with a different phase. The interference between
this ``simulacrum'' of the incident field and the portion of the
field which passes through the medium without being scattered
yields \emph{both} its phase delay in traversing the medium (thus
fixing its refractive index) and, adding to the non-radiative
power loss due to absorption, the power reduction of the
transmitted field. For what concerns power loss, this is
explicitly treated by the so-called ``Optical Theorem'' (OT), a
general and extremely useful result that holds true not only for
electromagnetic radiation but also for matter
waves~\cite{Newton1976}. Consider the simple case of a plane wave
with wave-vector $\textbf{k}_i = (\omega/c)\,\mathbf{\hat{k}}_i$,
where $\mathbf{\hat{k}}_i$ is a unit vector specifying the
incident direction, and polarized with the electric field along
$\mathbf{n}_i$, \mbox{$\mathbf{E}_i (\mathbf{r},t) =
\mathbf{n}_iE_i \exp[\I (\mathbf{k}_i \cdot \mathbf{r}-\omega t)
]$}, which encounters a scattering and absorbing medium confined
in a finite region of space around the origin. In far field, the
radiation scattered along $\hat{\mathbf{r}} =
\mathbf{r}/|\mathbf{r}|$ with wavevector $\textbf{k}_s = k_s
\hat{\mathbf{r}}$ can be written as
\begin{equation}\label{scatt_field}
    \mathbf{E}_s (\mathbf{r},t) = \mathbf{n}_s E_i\dfrac{\I S(\mathbf{k}_s,
    \mathbf{k}_i)}{kr}\E^{\I
(k_sr-\omega t)},
\end{equation}
where $\mathbf{n}_s$ is a vector normal to $\mathbf{k}_s$, which
depends on both $n_i$ and $k_s$, and $S(\mathbf{k}_s,
\mathbf{k}_i)$, called  the scattering amplitude, takes in general
complex values. Then, the OT states that the extinction cross
section
 is given by \footnote{The factor $k=2\pi/\lambda$
introduced the denominator of Eq.~(\ref{scatt_field}) makes
$S(\mathbf{k}_i, \mathbf{k}_s)$ dimensionless. By factoring out an
imaginary unit, which is related to the Gouy phase shift
accumulated in far-field between a spherical and a plane wave, the
amplitude function coincides in the short-wavelength limit (aside
from polarization effect) with a standard normalized diffraction
pattern~\cite{vandeHulst1968}. Note that in terms of the vector
scattering amplitude $\mathbf{f}= i (S/k) \mathbf{n}_s$ commonly
used in particle scattering, the optical theorem reads
$\sigma_{ext} = (4\pi/k^2) \mathrm{Im}[\mathbf{n}_i\cdot
\mathbf{f}(\mathbf{k}_s =\mathbf{k}_i)]$.}
\begin{equation}\label{Opt_Theo}
    \sigma_{ext} = \frac{4\pi}{k^2}(\mathbf{n}_i\cdot \mathbf{n}_s) \mathrm{Re}[S(\mathbf{k}_s
    =\mathbf{k}_i)]
\end{equation}
This rather surprising result, which basically shows that
evaluating the total extinction (due to scattering and, possibly,
absorption) of the radiation traversing the medium requires to
know the scattering amplitude \emph{only} in the forward direction
$\mathbf{k}_i$, can be rather easily obtained~\cite{Jackson1998}
by considering that, at steady state, the change in the total
energy density within a spherical region containing the whole
scattering medium is solely due to the dissipative absorption
processes taking place within it, and by equating the latter to
the incident plus scattered energy flow through the spherical
surface.\footnote{For optically anisotropic (birefringent) media,
the directions of polarization of the incident and scattered field
in the forward direction do not coincide, namely, the scattered
field contains a ``depolarized'' component that, being
perpendicular to $\mathbf{n}_i$ cannot of course interfere with
the incident beam. Nevertheless, Eq.~(\ref{Opt_Theo}) states that
the \emph{total} scattering cross section (including that due to
depolarized scattering) is still accounted for by the sole
polarized component. The reason for this correct, but apparently
paradoxical result is discussed in a recent
publication\cite{Degiorgio2009} concerning light scattering from
anisotropic colloidal particles.}

Most presentations of the OT consider an incoming plane wave and a
single scattering particle. Here we assume that the scattering
volume contains a large number of particles in Brownian motion.
The field scattered in any direction, except for  zero-scattering
angle, is the sum of many uncorrelated fields, it is a random
process with zero-average and a two-dimensional Gaussian
probability density. On the contrary, the fields scattered by
individual particles in the forward direction have all the same
phase, that is, the forward scattered wave exactly reproduces the
wavefront of the incident wave. In this paper we will only deal
with an incident plane wave, but the OT can be generalized to an
arbitrary incident field (for instance, a gaussian beam) by
considering the angular spectrum of the latter and applying
Eq.~(\ref{Opt_Theo}) to each plane wave component. Of course, in
such a case the relation between extinction and forward scattering
amplitude applies only to the \emph{whole} far-field diffraction
pattern.

The forward scattering amplitude contains \emph{more} information
than what simply conveyed by the OT. To see this, it is sufficient
to recall that, in a macroscopic description of the passage of
radiation though a material, the effects of propagation can be
fully embodied into a complex refractive index $\tilde{n} = n + i
n'$, whose real and imaginary parts are respectively related to
dispersion and power loss. Then, \emph{both} $n$ and $n'$ can be
formally linked to the real and imaginary parts of $S(\mathbf{k}_s
=\mathbf{k}_i)$. Yet, such a relation would be of little practical
interest unless we are able to evaluate $S(\mathbf{k}_s
=\mathbf{k}_i)$, which is the \emph{overall} forward scattering
amplitude, in terms of the specific microscopic scattering and
absorption events taking place in the medium. Strenuous efforts to
derive the macroscopic optical properties of a molecular fluid
from microscopic scattering events have spangled the history of
physical optics (for a review of the early attempts, see for
instance the books by Rosenfeld~\cite{Rosenfeld1951} and
Fabelinskii~\cite{Fabelinskii1968}), culminating in a series of
impressive contributions by Hynne and Bullough~\cite{Hynne1984,
Hynne1987, Hynne1990}, in which  a rigorous many-body
electrodynamic theory is used to obtain consistent expressions for
the refractive index, the extinction coefficient, and the
scattering cross section. Unfortunately, this powerful analysis,
which was performed with a very sophisticated formalism and
basically no approximation, leads to rather cumbersome general
results that, as a matter of fact, yield manageable expressions
only for rather dilute real gases, where interactions are
accounted for only at the level of the second virial coefficient
in a density expansion.

The situation looks however much more promising if we consider a
\emph{particulate} medium, namely, a collection of individual
scatterers, such as a suspension of colloidal particles dispersed
in a weakly scattering, non-absorbing solvent. Our problem can
then be rephrased as follows: is there any way to relate the real
and imaginary part of the refractive index of the whole dispersion
to the scattering properties of the \emph{individual} scatterers?
In the simple case of a thin slab of a medium consisting of a
dispersion of identical scatterers illuminated by a monochromatic
plane wave, the OT provides a straightforward affirmative answer
to this question, at least provided that two basic assumptions are
satisfied:~\cite{vandeHulst1968}
\begin{enumerate}
    \item The scatterers are \emph{randomly arranged}, namely, they do
    not display any structural correlation. This
    implies that the physical particles acting as scatterers interact
    very weakly, so that any correlations in density fluctuations
    can be neglected.
    \item The incident field ``seen'' by a particle coincides with the
    \emph{external} radiation, namely, any additional contribution due to the
    surrounding scatterers is neglected. Provided that we carefully
    specify that these contributions may be due not only to radiating, but also
    to quasi-static fields in the near-zone, which would actually be the case for
    those scatterers that lying a distance $r\lesssim \lambda$
    from the particle, this loosely means that ``multiple scattering'' effects are
    negligible.
\end{enumerate}
In this case, assuming for simplicity that the scatterers are
optically isotropic, and indicating with $\theta$ the polar angle
with respect to the direction of $\mathbf{k}_i$,
Eq.~(\ref{Opt_Theo}) reduces to
\begin{equation}\label{OT2}
   \sigma_{ext} = \frac{4\pi N}{k^2}\mathrm{Re}[s(0)],
\end{equation}
where $N=\rho V$ is the particle number in the volume $V$, $k =
2\pi/\lambda$, and $s(0)$ is the amplitude of the field scattered
in the forward direction $\theta =0$ by each \emph{single}
particle. This simpler relation can be obtained by considering
that i) for independent particles, $\sigma_{ext}$ is the sum of
the single-particle cross sections; ii) in the forward direction
(and \emph{only} in this direction) the scattering amplitudes  are
additive too, because all scattering contributions add \emph{in
phase}, regardless of the positions of the particles in the
scattering volume. By evaluating the phase shift in propagation
through the slab, the real and imaginary parts of the effective
refractive index of the medium are then easily found to
be\footnote{Because of the choice we made for the phase of the
incident and scattered field, the scattering amplitude we define
is the complex conjugate of the one defined by van de
Hulst\cite{vandeHulst1968}}
\begin{equation}\label{n_noninteracting}
    \left\{\begin{array}{l}
             n = 1- \dfrac{2\pi\rho}{k^3}\mathrm{Im}[s(0)]\vspace{3pt}\\
             n' = \dfrac{2\pi\rho}{k^3}\mathrm{Re}[s(0)] \\
           \end{array}\right.
\end{equation}
It is useful to recall that, when $n'\ne 0$, the intensity of a
plane wave propagating in the medium along $z$ decreases as $I
=I_0 \exp(-\gamma z)$, where the extinction coefficient
\begin{equation}\label{ext_coeff}
    \gamma = 2kn' = \frac{4\pi \rho}{k^2}\,\mathrm{Re}[s(0)] = \frac{\sigma_{ext}}{V}
\end{equation}
is simply the extinction cross section per unit volume.

Eq.~\ref{n_noninteracting} has been used to investigate the
effects of particle size on refractive index and extinction by
using the expression for scattering amplitude obtained from the
general Mie theory for light scattering from non-interacting
spherical particles.\cite{Champion1979}. However, repulsive and
attractive interparticle forces are well known to strongly affect
(the former by increasing, the latter by reducing) the
transmittance of light through a colloid. Moreover,
Eq.~(\ref{n_noninteracting}) suggests that also the real part $n$
of the refractive index should not be immune from interaction
effects. It is then very tempting to scrutinize whether an
effective refractive index could be defined in the interacting
case too, provided that the expression for the single-particle
forward scattering amplitude is suitably revisited to account for
the structure of the medium.

The goal of this work is to extend the OT approach to the case of
interacting colloidal particles, and to apply our results to
investigate the contribution of correlations to the refractive
index of a suspension. We shall confine our investigation to
suspensions of spherical particles in the colloidal size range
that, though possibly concentrated in terms of particle volume
fraction $\phi$, are still dilute in terms of number density $\rho
= \phi/v$, where $v$ is the volume of a single particle. This
restriction allows to describe interparticle forces using simple
model pair potentials and correlation functions, an approach which
is generally unsuited to properly describe the structure of dense
molecular fluids. In addition, we shall systematically adopt the
lowest order approximation for the optical properties of the
scatterers in which interactions effects do nevertheless show up
which, as we shall see, amounts to a second order approximation in
the optical polarizability $\alpha$. Even within these
approximations, however, an explicit evaluation of $S(0)$ casts
new light on the optical mechanism leading to the formation of the
transmitted wavefront, and highlights a rather unexpected role
played by multiple scattering, usually just regarded as a nuisance
in light scattering studies. A key result of our investigation is
indeed that, when considering the radiation strictly scattered in
the \emph{forward} direction, the contribution from multiple
scattering events, even when negligible at finite $q$, is
conversely found to be crucial to figure out why the forward
scattering amplitude, and therefore light extinction, depends on
interparticle interactions.

This paper is then organized as follows. The microscopic approach
we use and the approximations we make are introduced in
Section~\ref{s.dipoles}, where we first evaluate the electric
field in the forward direction due to the superposition of the
incident field with the waves generated by \emph{point-like}
oscillating particles. By considering a slab geometry, we show
that a microscopic expression for the refractive index of a
suspension of uncorrelated point-like particles is fully
consistent \emph{at all orders} with the high frequency limit of
the Clausius-Mossotti (CM) formula, namely, the Lorentz-Lorenz
expression\cite{Born1980}
\begin{equation}\label{Lorentz-Lorenz}
  \frac{n^2-1}{n^2+2}= \frac{4\pi}{3} \rho \alpha,
\end{equation}
provided that the expression for the dipole polarizability
includes the contribution from the reaction radiation field,
namely, the self-action of the dipole on itself.

The correlation contribution to the scattering amplitude and to
the refractive index for the general case of a homogeneous but
\emph{correlated} distribution of point-like dipoles is derived in
Section~\ref{corre}, and used to check that the forward scattering
amplitude is rigorously linked to $\sigma_{ext}$ by the Optical
Theorem. Such an explicit comparison yields an interesting
conceptual consequence: particle spatial correlations are
``encoded'' into the forward scattering amplitude only via the
additional contribution to the incident field brought in by the
secondary fields scattered by those dipole lying within a close-by
region with a size comparable to the correlation range of the
medium. In particular, we discuss the limits of validity of the CM
approximation in terms of the ratio of the correlation length
$\xi$ of the system to the wavelength $\lambda$ of the incident
light.

In Section~\ref{colloid}, we first extend the former results to a
system of particles of finite size in vacuum, comparing in
particular the limits for small and large particle size both in
the absence (\ref{Mie}) and in the presence (\ref{interacting}) of
interparticle interactions. Extension of these results to the case
of particles dispersed in a solvent is made in
Section~\ref{solvent}, where we show that this is straightforward
provided that the latter is assumed to be an uncorrelated
dielectric medium.  In Section~\ref{EffMedium} we frame our
results within the context of effective medium theories, which is
made, showing in particular that a very general result for the
effective \emph{static} dielectric constant, which is exact at 2nd
order in polarizability whatever the structural correlations of
the suspension, fails in the optical regime  when $\lambda \ll
\xi$, and has to be substituted by a novel, equally general
expression for the effective refractive index. Illustrative
examples of the contribution of interparticle interactions to the
concentration dependence of the refractive index are presented in
Section~\ref{HS} for the specific case of  hard-sphere
interactions. In particular, by investigating the strongly
correlated case of a colloidal fluid in equilibrium with a
colloidal crystal, we show that, whenever the peak of the
structure factor $S(q)$ falls within the detectable $q$-range, the
refractive index displays a peculiar ``anomalous dispersion''
region where it behaves similarly to the refractive index of a
Lorentz oscillator close to resonance. Experimental conditions in
which these effects could be observed are finally discussed in
Section~\ref{exper}.

\section{System of point-like particles}
\label{s.dipoles} The purpose of this Section is to describe the
total field scattered by a system of point-like polarizable
particles (namely, simple dipoles) by explicitly taking into
account the contribution to the incident field on each single
dipole due both the other surrounding dipoles \emph{and} to the
self-action of the dipole on itself. Consider then a collection of
$N$ oscillating dipoles made of a mobile charge $e$ and a fixed
charge $-e$ located at fixed positions $\bR_i$, with a spatial
distribution to be specified later. Defining the instantaneous
dipole moment of a given particle as
\begin{equation}
\bp(t) = \bp_0\E^{-\I\omega t},
\end{equation}
the electric field in $\br$ generated by the oscillating dipole
placed at the origin has the form~\cite{Jackson1998}
\begin{equation}
E_d^\mu(\br,t) = \E^{\I(kr-\omega t)} k^3 \Gamma^{\mu\nu}(\br)
\,p_0^\nu,
\end{equation}
where $k=\omega/c$, the radial unit vector is $n^\mu = r^{\mu}/r$,
and the dimensionless matrix $\Gamma(\br)$ is defined by
\begin{equation}
\Gamma^{\mu\nu}(\br) = \left (3 n^\mu n^\nu -\delta^{\mu\nu}\right
) \left [\frac{1}{(kr)^3}-\I \, \frac{1 }{(kr)^2}  \right ]-
\frac{1}{kr} \left (n^\mu n^\nu -\delta^{\mu\nu}\right ).
\label{gammamunu}
\end{equation}
Here and in the following, Greek superscripts refer to the spatial
components $(x,y,z)$ and the summation over repeated Greek indices
is understood. It is useful to observe right from the start that
the 2nd and 3rd term in $\Gamma^{\mu \nu}(\textbf{r})$, which
respectively account for the field in the so-called
``intermediate'' and ``radiation'' zones,\cite{Jackson1998} are of
order $r/\lambda$ and $(r/\lambda)^2$ with respect to the
electrostatic part decaying as $r^{-3}$. When we take into account
spatial correlations, the relative contribution of these two terms
will be found to increase with the correlation length $\xi$ of the
system.

If we include the presence of an external linearly polarized plane
wave of the form
\begin{equation}
E_0^\mu(\br,t) = \mathscr{E}^\mu  \E^{\I(\bk\cdot \br-\omega t)}
\label{ext}
\end{equation}
with $k^\mu \mathscr{E}^\mu =0$, the total electric field, due to
the external source and the collection of dipoles, is then
\begin{eqnarray}
E^\mu(\br,t) &=& E_0^\mu (\br,t) +\sum_j E_d^\mu(\br-\bR_j,t) \\
&=& E_0^\mu (\br,t) +k^3\sum_j \E^{\I k|\br-\bR_j|}
\Gamma^{\mu\nu}(\br-\bR_j) \, p_{0j}^\nu  \,\E^{-\I\omega t}
\label{eq1}
\end{eqnarray}
where $\bE_d(\br-\bR_j,t)$ is the contribution to the electric
field in $\br$ due to the dipole in $\bR_j$. Now we introduce the
polarizability $\alpha$ by assuming that the moment $\bp_j$ of the
dipole in $\bR_j$ is proportional to the local electric field due
to the {\it other} charges (i.e. the dipoles represent polarizable
point-like objects):
\begin{equation}
\bp_j(t) = \bp_{0j}\, e^{-\I\omega t} = \alpha\,\bE_j(\bR_j,t)
\end{equation}
where the subscript $j$  in $\bE_j(\bR_j,t)$ means that the
contribution due to the dipole in $\bR_j$ has to be subtracted,
and  $\alpha$ depends in general on the frequency $\omega$.

Yet, as we already mentioned, a consistent treatment requires  to
take also into account the action on the oscillating charge of the
field emitted by itself, namely, of the so-called ``radiation
reaction field'' which, in the absence of non-radiative
dissipation, provides the only mechanism for power loss. The
 question of the back-reaction of the radiated field onto
the motion of a charge is one of the most challenging problem in
electrodynamics, since it leads to an equation, originally derived
by Lorentz\cite{Lorentz1915} and then generalized to the
relativistic case by Abraham and Dirac\cite{Dirac1938}, which,
containing the derivative of $\ddot{\mathbf{r}}$, causes serious
difficulties due the appearance of ``runaway'' solutions showing
an exponential increase of $\ddot{\mathbf{r}}$ even in the absence
of external fields. Nevertheless, for a charge \emph{oscillating}
at non-relativistic speed, the ingenious approach devised by
Lorentz safely allows to include radiation reaction effects at
lowest order by introducing an imaginary additive contribution to
the particle polarizability\cite{vandeHulst1968}
\begin{equation}
\alpha \to \tilde\alpha = \alpha\,\left [ 1 + \I\,\frac{2}{3}
\,k^3 \alpha\right ] \label{alpha}
\end{equation}
In the following, the physical counterpart of the elementary
dipoles we introduced will be atoms or molecules excited at a
frequency $\omega$ far from any electronic or vibrational
transition, so that $\alpha$ will be taken as a \emph{real}
quantity.

Since we are considering point-like dipoles, the only intrinsic
length scale in the problem is the wavelength $\lambda$ of the
incident radiation. It is then suitable to define a dimensionless
polarizability $\alpha_d = \tilde\alpha k^3$ (where the subscript
$d$ stands for ``dipoles'') that, substituted into Eq.~(\ref{eq1})
using~(\ref{alpha}), yields:
\begin{equation}
E^\mu(\br,t) = E_0^\mu (\br,t) + \alpha_d \sum_j \E^{\I
k|\br-\bR_j|}  \Gamma^{\mu\nu}(\br-\bR_j) \, E_{j}^\nu(\bR_j,t)
\label{eq2}
\end{equation}
This is an equation for the electric field $\bE(\br,t)$, which can
be solved by iteration. To second order in the scaled
polarizability $\alpha_d$ the explicit solution is:
\begin{eqnarray}
E^\mu(\br,t) &=& E_0^\mu (\br,t) + \alpha_d \sum_j \E^{-\I
k|\br-\bR_j|} \,
\Gamma^{\mu\nu}(\br-\bR_j) \, E_0^\nu (\bR_j,t) + \nonumber \\
&&  \alpha_d^2 \sum_{j\ne l} \E^{\I k|\br-\bR_j|} \E^{\I
k|\bR_j-\bR_l| } \,
\Gamma^{\mu\nu}(\br-\bR_j)\Gamma^{\nu\sigma}(\bR_j-\bR_l)
E_0^\sigma (\bR_l,t) \label{eq3}
\end{eqnarray}

The second order approximation in $\alpha_d$ given by
Eq.~(\ref{eq3}) will be particularly useful in what follows both
to describe a system of interacting dipoles, and to extend our
results to the case of finite-size particles in
Section~\ref{colloid}, where, in the case of non-interacting
colloids, we shall also check for consistency with the exact Mie
results obtained within a continuum approach. However, it is
interesting to point out that Eq.~(\ref{eq2}) is also the starting
point of a \emph{non-perturbative} investigation of the dispersion
relation which characterizes the medium, which fully justifies the
CM relation for a system of uncorrelated dipoles, and actually
generalized it to account for the extinction contribution brought
in by radiation reaction effects. For a monochromatic
perturbation, the time dependence is factorized as
\begin{equation}
E_j^\mu(\br,t) = E_j^\mu(\br)\E^{-\I\omega t}
\end{equation}
Moreover, by evaluating the electric field at the position of the $i^{th}$ particle and subtracting the
singular contribution bue to the $i^{th}$ dipole, we get:
\begin{equation}
E_i^\mu(\bR_i) = E_0^\mu (\bR_i) + \bar\alpha \sum_{j\ne i} \E^{\I
k|\bR_i-\bR_j|} \, \Gamma^{\mu\nu}(\bR_i-\bR_j) \,
E_{j}^\nu(\bR_j) \label{mat1}
\end{equation}
For given positions of the $N$ dipoles of the medium
$(\bR_1\cdots\bR_N)$, this is a set of linear equations for the
$3N$ unknowns $E_i^\mu(\bR_i)$. Here we want to analyze the
possible solutions in the bulk, i.e., the monochromatic waves
which can propagate in the medium. Consider then a planar slab of
thickness $h$, placed orthogonally to the direction of propagation
$z$ of an incident plane wave polarized along $x$, $E_0^x(\br) =
\epsilon_0 \E^{\I kz} $. The field \emph{inside} the slab at the
position $Z_i$ along the optical axis of the $i$-th particle,
averaged on the positions of all the other particles, can always
be written as the superposition of two counter-propagating
components (the transmitted and the reflected wave)
\begin{equation}
E_i^x(\bR_i) = \epsilon^{+}\E^{\I qZ_i} + \epsilon^{-}\E^{-\I
qZ_i},
\end{equation}
where $q$ is a complex quantity to be specified later representing
the average wave vector of the propagating field inside the
medium. By substituting this parametrization into Eq. (\ref{mat1})
and introducing the pair distribution function $g(\br)$, we obtain
\begin{equation}
\epsilon^+ \E^{\I qz} + \epsilon^{-} \E^{-\I qz} = \epsilon_0
\E^{\I kz} + \bar\alpha \rho \int \D \br^\prime
\,g(\br-\br^\prime)\E^{\I k|\br-\br^\prime|}
\Gamma^{xx}(\br-\br^\prime) \, \left [ \epsilon^{+}\E^{\I
qz^\prime} + \epsilon^{-}\E^{-\I qz^\prime}\right] \label{mat3}
\end{equation}
where $\rho$ is the number density of particles in the system. We
anticipate that the adopted procedure is in fact correct to second
order in an expansion in powers of the molecular polarization,
while it neglects three body correlations.

The domain of integration in Eq. (\ref{mat3}) coincides with the
whole volume of the slab. Note that $g(0)=0$ due to the presence
of a hard core: this guarantees that the constraint $j\ne i$ in
Eq. (\ref{mat1}) is correctly implemented. Next we write $g(\br) =
1 + h(r)$, where $h(r)$ is non-zero only at short range (i.e. only
for $r$ comparable to the molecular diameter). The first
contribution ($g(\br)=1$) accounts for the average particle
distribution, while the residual term, containing $h(r)$, provides
the correlation contribution to the propagating wave. Let us
examine the uncorrelated term with the supplementary {\it caveat}
to exclude an infinitesimal neighborhood of $\br=0$. By explicitly
performing the integrals we obtain the following set of four
consistency conditions:
\begin{eqnarray}
 \frac{\epsilon^+}{k-q} +\frac{\epsilon^-}{k+q}
 &=& -\frac{k^2\epsilon_0}{2\pi\bar\alpha\rho } \label{disp1}\\
\frac{\epsilon^+\E^{-\I(k+q)h}}{k+q} +
\frac{\epsilon^-\E^{\I(k-q)h}}{k-q}  &=& 0 \label{disp2}\\
 \frac{\epsilon^\pm}{k+q} +\frac{\epsilon^\pm}{k-q} -
\frac{2}{3} \,\frac{\epsilon^\pm}{k} &=& -\frac{k^2\epsilon^\pm
}{2\pi\bar\alpha\rho} \label{disp3}
\end{eqnarray}
which are necessary and sufficient for the validity of
Eq.~(\ref{mat3}) for all $z$. The condition expressed by
Eq.~(\ref{disp1}) is equivalent to the extinction theorem: the
incident wave of wave-vector $k$ does not propagate in the medium
because it is exactly canceled by the contribution of the
oscillating dipoles. Moreover, together with Eq.~(\ref{disp2}), it
provides the amplitudes of the waves propagating in the direction
of the incident signal ($\epsilon^+$) and in the opposite
direction ($\epsilon^-$). Finally, the last two
equations~(\ref{disp3}) allow to fix the wave-vector $q$ of the
wave at frequency $\omega=kc$ propagating inside the medium. By
defining the complex refractive index $\tilde{n}$ via
$q=\tilde{n}k$, the final result is:
\begin{equation}
\frac{\tilde{n}^2-1}{\tilde{n}^2+2} = \frac{4\pi}{3}\,\alpha_d
\rho_d= \frac{4\pi}{3}\,\tilde\alpha \rho \label{ll}
\end{equation}
where $\rho_d =k^{-3}\rho$ is a dimensionless dipole density.
Notably, Eq.~(\ref{ll}) is a generalized Clausius--Mossotti (or,
better, Lorentz-Lorenz) relation for the \emph{complex} refractive
index $\tilde{n}$, which includes the effects of radiation
reaction through the imaginary part of $\tilde\alpha$. Expanding
$\tilde n$ at second order in $\tilde\alpha \rho$,
using~(\ref{alpha}) with $\alpha$ real, and equating the real and
imaginary parts, we obtain
\begin{equation}
\label{refr_index_uncorrelated}
    \left\{\begin{array}{l}
             n =  1 + 2\pi\alpha\rho
+\dfrac{2}{3}\,(\pi\alpha\rho)^2\vspace{3pt}\\
             n' = \dfrac{4\pi}{3}\rho k^3\alpha^2\\
           \end{array} \right.
\end{equation}
The real part coincides with the expansion at 2nd order of the
usual CM formula, whereas the dissipative radiation-reaction term
contribute \emph{only} to attenuation. When this result is
inserted into Eq.~(\ref{n_noninteracting}), it yields an explicit
expression for the scattering amplitude of a \emph{single}
non-interacting dipole
\begin{equation}\label{cross_section_dipoles}
    s_0(0) = \frac{2}{3}k^6\alpha^2 - \I k^3\alpha,
\end{equation}
to leading order in $\alpha$ (linear for the imaginary, and
quadratic for the real part). From the OT we then get the correct
extinction cross section $\sigma_{ext} = (8\pi/3)N \alpha^2 k^4$
for Rayleigh scattering from independent particles with a size
much smaller than $\lambda$.

\section{Correlated fluid of point-like dipoles}
\label{corre} We now consider a correlated dielectric medium
starting from the general expression (\ref{mat3}). The weight
function $h(\br-\br^\prime)$ is appreciably different from zero
only in a small neighborhood of $\br$, therefore if the
observation point $\br$ is placed in the bulk, we can extend the
integral to the whole space, neglecting the effects of the
boundary surfaces. The resulting consistency condition, which
corrects the Lorentz-Lorentz formula (\ref{ll}) for a correlated
fluid, is obtained by including into Eq.
(\ref{disp1}--\ref{disp3}) a correlation integral $C(q,k)$:
\begin{eqnarray}
\label{eqn}
1 &=& 4\pi\tilde\alpha\rho \,\left [ \frac{1}{\tilde n^2-1} +\frac{1}{3} + C(q,k) \right] \\
C(q,k) &=& \frac{k^3}{4\pi} \, \int \D\br \,\E^{-\I
qz}h(r)\Gamma^{xx}(\br)\E^{\I kr} \label{ci}
\end{eqnarray}
where, as usual, an infinitesimal neighborhood of $r=0$ is
excluded from the integration domain. By introducing the Fourier
transforms, the correlation integral can be expressed as
\begin{equation}
C(q,k)= \frac{1}{2q^2}\,\int \frac{\D\bp}{(2\pi)^3} \, h(p)\,\left
[ \frac{k^2q^2+(q^2-\bp\cdot\bq)^2} {|\bq-\bp|^2-k^2+\I\eta}
-\frac{q^2}{3}\right ] \label{intc}
\end{equation}
where the complex wavevector $\bq=\tilde n\bk$ is directed along
$z$. Eq.~(\ref{eqn}), with the definition~(\ref{intc}), implicitly
relates the complex refraction index $\tilde n$ to the microscopic
(complex) polarizability $\tilde\alpha$ in a correlated fluid of
number density $\rho$. To {\it second} order in the scaled
polarizability, the correlation integral (\ref{intc}) can be
evaluated at $\bq=\bk$. In this case, Equation (\ref{eqn})
explicitly provides $\tilde n$ as a function of $\tilde\alpha$
with the result:
\begin{equation}
\tilde{n}^2 = 1 +
\frac{4\pi\tilde\alpha\rho}{1-4\pi\tilde\alpha\rho\,[\frac{1}{3}+C(k,k)]}
\end{equation}
which reduces to (\ref{ll}) for $C=0$ and represents an
approximate, non perturbative expression of the complex refractive
index of a correlated medium. Expanding again to second order, we
obtain the exact lowest order correction to the refractive index
in a correlated fluid:
\begin{equation}
\tilde{n} = 1 + 2\pi\tilde\alpha\rho
+\frac{2}{3}\,(\pi\tilde\alpha\rho)^2 + 8(\pi \tilde\alpha\rho)^2
C. \label{ennet}
\end{equation}
Recalling that $\tilde\alpha \rho =\alpha_d\rho_d$, we point out
that this expansion can be regarded as valid at second order in
$\alpha_d$ with \emph{no restriction} on the value of the scaled
density $\rho_d$ (namely, it is not a low-density expansion). The
formal expressions of the real and imaginary part of $C$ read:
\begin{eqnarray}
\label{reim} {\rm Im}\, C &=& \frac{1}{16\pi k^3}\int_0^{2k} \D
p\,p h(p) \, \left [ 2k^4 -k^2p^2+\frac{p^4}{4}\right]
\\
{\rm Re}\, C &=& \frac{1}{16\pi^2k^3}\,\int_0^\infty \D p\,p h(p)
\left [ \frac{8}{3}\,k^3p-kp^3+\left ( 2k^4
-k^2p^2+\frac{p^4}{4}\right)\ln\frac{p+2k}{|p-2k|} \right ]
\nonumber
\end{eqnarray}
Here and in the following we drop the momentum dependence of the
correlation integral, setting $C=C(k,k)$. To this order of
approximation, both the refractive index $\tilde{n}$ and the
forward scattering amplitude $S(0)$ acquire an \emph{additive}
contribution due to correlations:
\begin{equation}\label{essec}
    \left\{\begin{array}{l}
             \delta S(0) = -\I\tilde\alpha^2k^3VC \\
             \delta \tilde{n}= 2\pi \tilde\alpha^2C\\
           \end{array}\right.
\end{equation}
where $V$ is the volume of the sample. Notably, if we define an
excess scattering amplitude \emph{per particle}, $\delta s (0)=
\delta S(0)/N$, so that $s(0) = s_0(0)+\delta s(0)$,
Eq.~(\ref{n_noninteracting}) remains then formally valid, although
of course the effective forward scattering amplitude $s(0)$
actually depends on $\rho$ and on the specific structure of the
medium via the correlation integral $C$.

It is useful to point out that the correlation contribution to the
imaginary part $n'$ of the refractive index in Eq.~(\ref{reim})
depends \emph{only} on those values of $p$ that are smaller than
the maximum wave-vector $2k$ (corresponding to a scattering angle
$\theta = \pi$) falling within the experimentally detectable
range. Although this is seemingly not the case for
$\mathrm{Re}\,C$, we shall see in Section~\ref{HS} that the actual
occurrence or not of a peak of the structure factor $S(q)$ within
the accessible range $q \le 2k$ \emph{does} appreciably influence
the value of the refractive index $n$. Notice also that
Eq.~(\ref{ennet}) provides a quantitative explanation of the
reason why the Lorentz-Lorenz expression for the refractive index
of a \emph{molecular} fluid is often a very good approximation,
even in the presence of consistent correlations. For $k\xi \ll 1$,
where $\xi$ is the correlation length defined as the distance
where $h(r)$ becomes negligible,\footnote{Formally, $\xi$ can be
defined as
    $\xi = \int d\mathbf{r} r^2 h(r)/\int d\mathbf{r} r
    h(r)$,
which for an exponentially-decaying correlation function coincides
with the decay length.} the correlation coefficient $C$ is indeed
easily found to behave as $(k\xi)^2 \sim  (\xi/\lambda)^2$. Then,
provided that $\xi$ is of the order of the molecular size (which
is usually the case, unless the system is close  to a critical
point), correlation corrections are small.

Expression (\ref{essec}) can be readily shown to be fully
consistent with the Optical Theorem. We first evaluate the real
part of Eq.~(\ref{essec}) through Eq.~(\ref{reim}), retaining the
correlation contribution and radiation-reaction effects:
\begin{equation}
{\rm Re}[S(0)] = N\,\alpha^2\,\left \{ \frac{2}{3} k^6 +
\frac{\rho}{4}\, \,\int_0^{2k} \D q\,q h(q) \left [ 2k^4
-k^2q^2+\frac{q^4}{4}\right]\right \} \label{res1}
\end{equation}
where the first term comes from radiation reaction in the first
order contribution, while the second from Eq.~(\ref{reim}). It is
convenient to change the integration variable to
$q=2k\sin(\theta/2)$ with $\theta\in (0,\pi)$. Eq.~(\ref{res1})
then becomes
\begin{eqnarray}
{\rm Re}[S(0)] &=& N\alpha^2k^6\left \{ \frac{2}{3} +
\frac{\rho}{4}\int_0^{\pi} \D\theta\,\sin\theta(1+\cos^2\theta)h(q)\right \} \nonumber\\
&=& N\frac{\alpha^2k^6}{4}\int_0^{\pi}
\D\theta\,\sin\theta(1+\cos^2\theta)[1+\rho\, h(q)] \label{res2}
\end{eqnarray}
Calling $\theta$ the scattering angle and $\varphi$ is the angle
between the scattering plane and the polarization vector, the
differential cross section for Rayleigh scattering from a
collection of dipoles is given by \cite{Jackson1998}:
\begin{equation}
\frac{d\sigma}{d\Omega} = N\alpha^2\left(\frac{\omega}{c}\right)^4
(1-\sin^2\theta\cos^2\varphi)\, [1+\rho\, h(q)] \label{diff},
\end{equation}
where $q=2k\sin(\theta/2)$ and, for a harmonically bound
oscillator of elementary charge $e$ excited at a frequency
$\omega$ much lower than its natural frequency $\omega_0$, $\alpha
= -e^2/(m\omega_0^2)$. Putting again $s(0) = S(0)/N$, we
immediately verify via an integration of Eq.~(\ref{diff}) on the
solid angle $d\Omega =\sin\theta \,\D\theta\,\D\varphi$, namely,
by averaging over all possible orientations of the incident field
with respect to the scattering plane, that Eq.~(\ref{OT2}) is
satisfied by our final expression~(\ref{res2}).

\section{Colloidal Suspensions}
\label{colloid}
 Up to now we considered just point-like
polarizable particles, i.e. particles whose size is much smaller
than the wavelength of the incident field. However, if we are
interested in colloidal suspensions, we have to deal with
polarizable spheres whose size may be comparable to or even larger
than the optical wavelength.  To this aim, we  model each particle
$p$ as a homogeneous dielectric sphere of radius $a$ made of $M$
polarizable molecules. On a microscopic scale, the system is again
described by a collection of point-like dipoles, whose spatial
distribution clusters however into spherical units centered around
the position of the center of mass of each single colloidal
particle. The derivation of the previous Sections is therefore
still valid, provided the polarizability $\alpha$ is the
microscopic polarizability of each molecule, the density $\rho$ is
the number density of molecules, related to the colloidal particle
density $\rho_p$ by $\rho_p=\rho/M$ and the distribution function
$h(r)$ has a non-trivial structure, appropriate for the underlying
``cluster fluid".

Let us consider a collection of $N$ spherical particles,
characterized by a normalized probability distribution
$P_p(\bR_1\cdots \bR_N)$. Each polarizable molecule is identified
by its position $\br^l_m$, where $l=1\cdots N$ labels the colloid
and $m=1\cdots M$ the specific molecule in the colloid. If the
molecules are homogeneously distributed inside each sphere in an
uncorrelated way, their probability density in space is given
by:
\begin{equation}
P(\{\br^l_m\}) = \int \D\bR_1\cdots \D\bR_N\, P_p(\bR_1\cdots
\bR_N)\, \prod_{l,m} \frac{ \theta \left ( a-|\br^l_m
-\bR_l|\right )}{v} \label{prob}
\end{equation}
where $\theta(x)$ is the Heaviside step function and
$v=(4\pi/3)a^3$ is the particle volume. The molecular distribution
enters our expressions through the correlation integral Eq.
(\ref{ci}) where we used the standard definition of radial
distribution function:\cite{Hansen2013}
\begin{equation}
\rho^2\,g(\bR-\bR^\prime) = \left < \sum_{i\ne j}
\delta(\bR-\bR_i)\,\delta(\bR^\prime-\bR_j)\right >
\end{equation}
Now, this expression must be generalized to:
\begin{equation}
\rho^2\,g(\br-\br^\prime) = \left < \sum_{(l,m)\ne
(l^\prime,m^\prime)}
\delta(\br-\br^l_m)\,\delta(\br^\prime-\br^{l^\prime}_{m^\prime})\right
>
\end{equation}
where the average is taken according to the probability
distribution (\ref{prob}). In performing the average, we must
consider two possibilities in the summation over particle pairs:
\begin{itemize}

\item $l=l^\prime$ (and then $m\ne m^\prime$). These terms take into account
spatial correlations among molecules inside the same sphere,
induced by their confinement. The resulting contribution to
$\rho^2 g(\br-\br^\prime)$ is:
\begin{equation}
\rho\, M\, \frac{1}{v^2}\int \D\bR \,\theta \left ( a -|\br
-\bR|\right ) \theta \left (a-|\br^\prime -\bR|\right )
\end{equation}
The convolution integral is easily performed in Fourier space by
introducing the form factor
\begin{equation}
F(q) = \frac{1}{v} \int \D\bR \,\theta \left ( a-r\right
)\E^{\I\bq\cdot\bR} = \frac{j_1(qa)}{qa}, \label{form}
\end{equation}
where
$$j_1(x)  = 3\,\frac{\sin x -x\cos x}{x^3}$$ is the 1st order spherical Bessel function of the first kind.
\item $l \ne l^\prime$.
This term takes into account the correlations between molecules belonging
to different colloids. The resulting contribution is:
\begin{equation}
\rho^2\,\frac{1}{v^2}\,\int \D\bR
\D\bR^\prime\,g_p(\bR-\bR^\prime)\, \theta \left ( a -|\br
-\bR|\right ) \theta \left ( a- |\br^\prime -\bR^\prime|\right )
\end{equation}
where $g_p(r)$ is the distribution function of the colloidal
particles.
\end{itemize}
In summary, our final expressions for the correlation contribution
to the refractive index (\ref{reim}) are still valid with the
substitution
\begin{equation}
\rho^2\,h(q) \to  M^2\,\rho_p\,F(q)^2\,\left [ 1 + \rho_p
\,h_p(q)\right ]. \label{subs}
\end{equation}
We note two main differences with respect to the previous
expressions: $i)$ the presence of the form factor $F(q)^2$ and
$ii)$ the additive contribution (the unity in the square bracket).
The latter takes care of the scattering from pairs of molecules
inside the same colloid, which in turns provides the second order
contribution in the Mie scattering of each colloidal
particle.~\footnote{Notice that, as a matter of fact, $F(q)$ is
the purely ``geometrical'' form factor obtained for the scattering
from a uniform sphere in the Rayleigh--Gans approximation by
assuming that the incident field on each  volume element is the
\emph{unperturbed} external field} It is also important to notice
that we have in this case an \emph{intrinsic} structural length
scale, given by the particle size $a$. It is then suitable to
perform the expansion in terms of the particle polarization per
unit volume  $\alpha_p =M\alpha/v$, a dimensionless quantity that
plays the same role as $\alpha_d$ for point-like dipoles.
Substituting~(\ref{subs}) into the correlation integral
(\ref{ci}), we find at 2nd order in $\alpha_p$:
\begin{equation}
\tilde{n} = 1 + 2\pi\phi\, \alpha_p +2\pi
\left(\frac{\pi}{3}\phi^2+\widetilde{C}\phi\right)\alpha_p^2
\label{ncolle}
\end{equation}
where we have defined a dimensionless complex \emph{correlation
factor} $\widetilde{C} = C_r+\I C_i$, with:
\begin{eqnarray}
C_r &=& \frac{v}{4\pi k^3}\int_0^\infty \D q\,q F^{\,2}(q) \left [
1+ \rho_p h(q)\right]  \left [\frac{8}{3}k^3q-kq^3+\left(2k^4
-k^2q^2+\frac{q^4}{4}\right)
\ln\frac{q+2k}{|q-2k|} \right ] \nonumber \\
C_i &=&  \frac{v}{4k^3}\int_0^{2k} \D q\,q F^{\,2}(q) \left [ 1+
\rho_ph_p(q)\right ] \left [ 2k^4
-k^2q^2+\frac{q^4}{4}\right].\label{cc}
\end{eqnarray}
We stress again that Eq.~(\ref{ncolle}) is valid, at second order
in $\alpha_p/v$, for \emph{any} value of $\phi$.

For the real part $n$ of the refractive index and the extinction
coefficient $\gamma = 2kn'$, which are the experimentally observed
quantities, Eq.~(\ref{ncolle}) yields:
\begin{equation}\label{ncolle_parts}
    \left\{\begin{array}{l}
             n = 1 + 2\pi\phi \,\alpha_p +2\pi
\left(\dfrac{\pi}{3}\phi^2+C_r\phi\right)\alpha_p^2 \vspace{5pt}\\
             \gamma = 4\pi k \phi C_i\,\alpha_p^2\\
           \end{array} \right.
\end{equation}

 For an easier comparison with the experimental data,
and to check for consistency in the absence of interparticle
interactions with the continuum Mie theory, it is useful to
introduce the index of refraction $n_p$ of the material
constituting the colloidal particle. Expanding the CM equation
inside the particle at second order in the refractive index
contrast $\Delta n_p =n_p -1$, the particle polarizability per
unit volume is easily found to be given by
\begin{equation} \label{particlepol}
    \alpha_p =\frac{1}{4\pi}\left[2\Delta n_p
    -\frac{(\Delta n_p)^2}{3}\right].
\end{equation}
Retaining for consistency only terms to order $(\Delta n_p)^2$,
Eq. (\ref{ncolle_parts}) becomes:
\begin{equation}\label{ncolle_2}
\left\{\begin{array}{l}
         n = 1+\phi\,\Delta n_p  + \left[\dfrac{\phi-1}{3}+\dfrac{C_r}{\pi} \right]\dfrac{\phi}{2}\,(\Delta n_p)^2 \vspace{5pt}\\
         \gamma =  \dfrac{k\phi}{\pi}C_i\,(\Delta n_p)^2 = \dfrac{2\phi}{\lambda}C_i\,(\Delta n_p)^2\\
       \end{array} \right.
\end{equation}
Notice that for $C_r = 0$ the refractive index is given by:
\begin{equation}\label{n-totallyuncorr}
    n= 1+\left(1-\frac{\Delta n_p}{6}\right)\Delta n_p \phi
    +\mathcal{O}(\phi^2)
\end{equation}
which therefore differs at \emph{first} order in $\phi$, even in
the absence of both intra- and inter-particle correlations, from
the simple expression $n=1+\Delta n_p \phi$, obtained by
volume-averaging the refractive indices of particle and solvent
(which is conversely correct for polarizabilities).

For what follows, it is also useful to introduce, as customary in
light scattering theory, an ``efficiency factor'' $Q_{ext}$,
defined as the ratio of $\sigma_{ext}$ to the total geometric
cross--section $N\pi a^2$ of the particles.  Taking into account
the definition of $\gamma$ in~(\ref{ext_coeff}), we have:
\begin{equation}\label{efficiency1}
   Q_{ext} = \frac{\sigma_{ext}}{N\pi a^2} =
   \frac{4a}{3\phi}\,\gamma,
\end{equation}
so that, from the second of~(\ref{ncolle_2}):
\begin{equation}\label{efficiency2}
   Q_{ext} = \frac{\sigma_{ext}}{N\pi a^2} =
   \frac{4ka}{3\pi}\,C_i (\Delta n_p)^2.
\end{equation}

\subsection{Non-interacting particle limit and comparison with Mie theory}
\label{Mie} We first examine the limit, denoted by the superscript
``0'', in which inter-particle correlations can be neglected, that
is obtained by setting $h(q) \equiv0$ in Eq.~(\ref{cc}):
\begin{equation}\label{intracorr}
    \left\{\begin{array}{l}
             C_r^{\,0}  = \dfrac{1}{3
x^3}{\displaystyle \int_0^\infty } \D y \,y F^{\,2}(y)\left [
\dfrac{8}{3}x^3y-xy^3+\left (2x^4 -x^2y^2+\dfrac{y^4}{4}\right)\,
\ln\dfrac{y+2x}{|y-2x|} \right] \vspace{5pt} \\
             C_i^{\,0}   = \dfrac{\pi}{3x^3}{\displaystyle \int_0^{2x}} \D y\,y
F^{\,2}(y)\left [ 2x^4 -x^2y^2+\dfrac{y^4}{4}\right], \\
           \end{array} \right.
\end{equation}
where $x=ka$ and $y=qa$. This single-particle approximation will
be compared to the Mie solution, expanded at 2nd order in $\Delta
n_p$. It is worth considering the cases of particles much smaller
or much larger than the wavelength separately.
\paragraph{Small particles ($x\ll 1$)}
In the limit $x\to 0$, the real and imaginary parts of the
correlation factor in~(\ref{intracorr})  are easily found to be:
\begin{equation}\label{small ka}
    \left\{\begin{array}{l}
              C_r^{\,0} \underset{x\rightarrow 0}{\longrightarrow} \dfrac{88\pi}{75}\,x^2 \vspace{5pt}\\
              C_i^{\,0}\underset{x\rightarrow 0}{\longrightarrow} \dfrac{8\pi}{9}\,x^3\\
           \end{array}\right.
\end{equation}
Substituting in Eq. (\ref{ncolle_2}, \ref{efficiency2}), we find
the limiting behaviour:
\begin{equation}\label{n_small}
\left\{\begin{array}{l}
         n^0 \underset{x\rightarrow 0}{\longrightarrow}1+\phi\,\Delta n_p  - \dfrac{(\Delta n_p)^2}{6}\,\phi(1-\phi) + \dfrac{44}{75} (\Delta n_p)^2\phi \,x^2 \\
         \gamma^0 \underset{x\rightarrow 0}{\longrightarrow} \dfrac{8}{9}\,kx^3\phi\,(\Delta n_p)^2\vspace{5pt}\\
         Q_{ext}^{\,0} \underset{x\rightarrow 0}{\longrightarrow} \dfrac{32}{27}\,x^4\,(\Delta n_p)^2\\
       \end{array}\right.
\end{equation}
where in the first equation we have also retained the lowest-order
dependence on $x$, for later convenience. Reassuringly,
$Q_{ext}^{\,0}$ coincides with the efficiency factor for Rayleigh
scatterers, namely, for particles much smaller than the
wavelength.\cite{vandeHulst1968} It is also very interesting to
notice that, using Eq.~(\ref{n_noninteracting}) the real part of
the scattering amplitude can be written
\begin{equation}\label{rr_sphere}
    {\rm Re}\, s^0(0) = \frac{8}{27}(\Delta n_p)^2 x^6 = (2/3) \,
    (v\alpha_p)^2\,k^6,
\end{equation}
which, comparing with Eq.~(\ref{cross_section_dipoles}), is
identical to the radiation reaction contribution from a single,
\emph{point-like} dipole of polarizability $v\alpha_p$. This
result is equivalent to the brilliant conclusion reached by
Lorentz: the radiation reaction from a spherical radiator with
fixed polarizability does not depend on its size, provided that
the latter is much smaller than the wavelength. It also clarifies,
however, a subtle feature of the general results obtained in the
former Section. In deriving Eq. (\ref{ncolle}), we have actually
disregarded the radiation reaction term of each polarizable
molecule because, due to the presence of the dimensionless factor
$\alpha k^3 \ll 1$ in Eq. (\ref{alpha}), this gives a negligible
contribution to the scattering of the whole colloidal particle.
Surprisingly, therefore, while the extinction from a distribution
of uncorrelated point-like dipoles is \emph{solely} due to
radiation reaction, when the same dipoles ``cluster'' into uniform
spherical particles this contribution becomes vanishingly small.
The microscopic approach we followed shows that it is again
multiple scattering (in the generalized sense stated in
Section~\ref{intro}) that, due to internal correlations, generates
a ``collective'' radiation reaction effect, leading to finite
extinction.

\paragraph{Large particles ($x\gg 1$).}
In the opposite case $x\to\infty$, the real and imaginary part of
the correlation factor in Eq.~(\ref{intracorr}) can be readily
evaluated at leading order in $x$ in terms of simple integrals of
$j_1(y)$ with the result:
\begin{equation}\label{C_p large x}
    \left\{\begin{array}{l}
             C_r^{\,0} \underset{x\to\infty}{\longrightarrow} \dfrac{7\pi}{3} \vspace{5pt}\\
             C_i^{\,0} \underset{x\to\infty}{\longrightarrow}\dfrac{3\pi}{2}x \\
           \end{array}\right.
\end{equation}
which, using Eq.~(\ref{ncolle_2}), yields
\begin{equation}\label{Mie_large}
    \left\{\begin{array}{l}
             n^0 \underset{x\rightarrow \infty}{\longrightarrow} 1+\phi\Delta n_p  +\phi\left(1+\dfrac{\phi}{6}\right)(\Delta n_p)^2 \vspace{5pt}\\
             \gamma^0 \underset{x\rightarrow \infty}{\longrightarrow} \dfrac{3\phi}{2a}\,x^2 (\Delta n_p)^2 =
             2\pi\rho a^4 k^2 (\Delta n_p)^2 \vspace{5pt}\\
             Q_{ext}^{\,0} \underset{x\rightarrow \infty}{\longrightarrow} 2x^2(\Delta n_p)^2
           \end{array}\right.
\end{equation}
The expression for $\gamma^0$ and $Q_{ext}^{\,0}$
in~(\ref{Mie_large}) are however quite suspicious: in fact, they
highlight a severe limit in the quadratic expansion we use.
Indeed, from~(\ref{Mie_large}) $Q_{ext}^{\,0}$ grows without
limits with $x$, whereas in Mie theory \mbox{$Q_{ext}^{\,0}
\underset{x\rightarrow \infty}{\longrightarrow} 2$}, whatever the
value (even complex) of the particle refractive
index.\footnote{This is known as the ``extinction paradox'', since
a very large particle apparently ``casts a shadow'' which is the
double of its geometrical cross section. Consistency with the
``macroscopic'' experience is recovered by noticing that half of
the scattered light is scattered in an extremely narrow
diffraction cone around the forward direction, which could be
excluded only using a detector with an acceptance angle $\vartheta
\ll \lambda /2a$} Actually, the efficiency factor
in~(\ref{Mie_large}) coincides with the value obtained in the
Rayleigh--Gans (RG) approximation of the exact Mie solution, which
requires both $\Delta n_p$ \emph{and} the maximum phase delay
$\delta =2x\Delta n_p$ that the incident field undergoes in
traversing the particle to be small.\cite{vandeHulst1968} The
second condition, in particular, is equivalent to assume that the
incident radiation on each volume element of the particle
coincides with the external field. In our description, this means
that, for $\delta \ll 1$, internal multiple scattering
contributions are negligible, so that intra-particle correlations
are only related to the geometrical arrangement of the elementary
scatterers expressed by the form factor. Moreover, since for large
particles $Q_{ext}^{\,0} \simeq \delta^2 /2$, the RGD condition is
met only when $Q_{ext}^{\,0}\ll 1$, namely, when the extinction
cross-section is substantially smaller than the geometrical
``shadow'' of the particle.

As a matter of fact, in the double limit $x\rightarrow \infty$,
$\Delta n_p \rightarrow 0$, made by keeping $\delta$ finite, known
in the light scattering jargon as the ``anomalous diffraction''
limit, it is possible to find an \emph{exact} solution for $s(0)$,
given in our notation by (see Section 11.22 in van de
Hulst\cite{vandeHulst1968}):
\begin{equation}\label{Anodiff}
   s (0) = x^2\left(\frac{1}{2} +\I \,\frac{\E^{\I \delta}}{\delta} +\frac{1-\E^{\I \delta}}{\delta^2}  \right),
\end{equation}
which yields, for the efficiency factor:
\begin{equation}\label{efficiency-Anodiff}
   Q_{ext}^{\,0} (\delta)= \frac{4\phantom{^2}}{x^2}\,\mathrm{Re}\,s(0) = 2 - \frac{4}{\delta}\sin(\delta) + \frac{4\phantom{^2}}{\delta^2}(1-\cos \delta),
\end{equation}
Whereas $ Q_{ext}^{\,0} (\delta)\underset{\delta \rightarrow
0}{\longrightarrow} \delta^2/2$, for $\delta \gg 1$ the scattering
cross section per particle $\pi a^2 Q_{ext}^{\,0}$ correctly
converges to twice the geometrical shadow. This finite limiting
value, which does \emph{not} depend on $\Delta n_p$\footnote{Of
course, if $\Delta n_p$ is \textit{identically} equal to 0,
$Q_{ext}^{\,0}$ must vanish. What (\ref{efficiency-Anodiff})
actually means is that $Q_{ext}^{\,0}(\delta)$ is discontinuous:
$\lim_{\delta\rightarrow 0}Q_{ext}^{\,0}(\delta) \ne Q_{ext}^{\,0}
(0)$} and corresponds to the limit of diffraction optics, can be
recovered only by resumming \emph{all} orders in $\Delta n_p$,
however small they are, and is therefore missing in our analysis.
Technically, this is due to the fact that the Mie solution,
expressed as a series depending on the two parameters $n_p$ and
$x$, is not absolutely convergent, therefore, exchanging the
limits $x\rightarrow \infty$ and $n_p \rightarrow 1$ is therefore
not permitted.

\begin{figure}[h!]
\includegraphics[width=\textwidth]{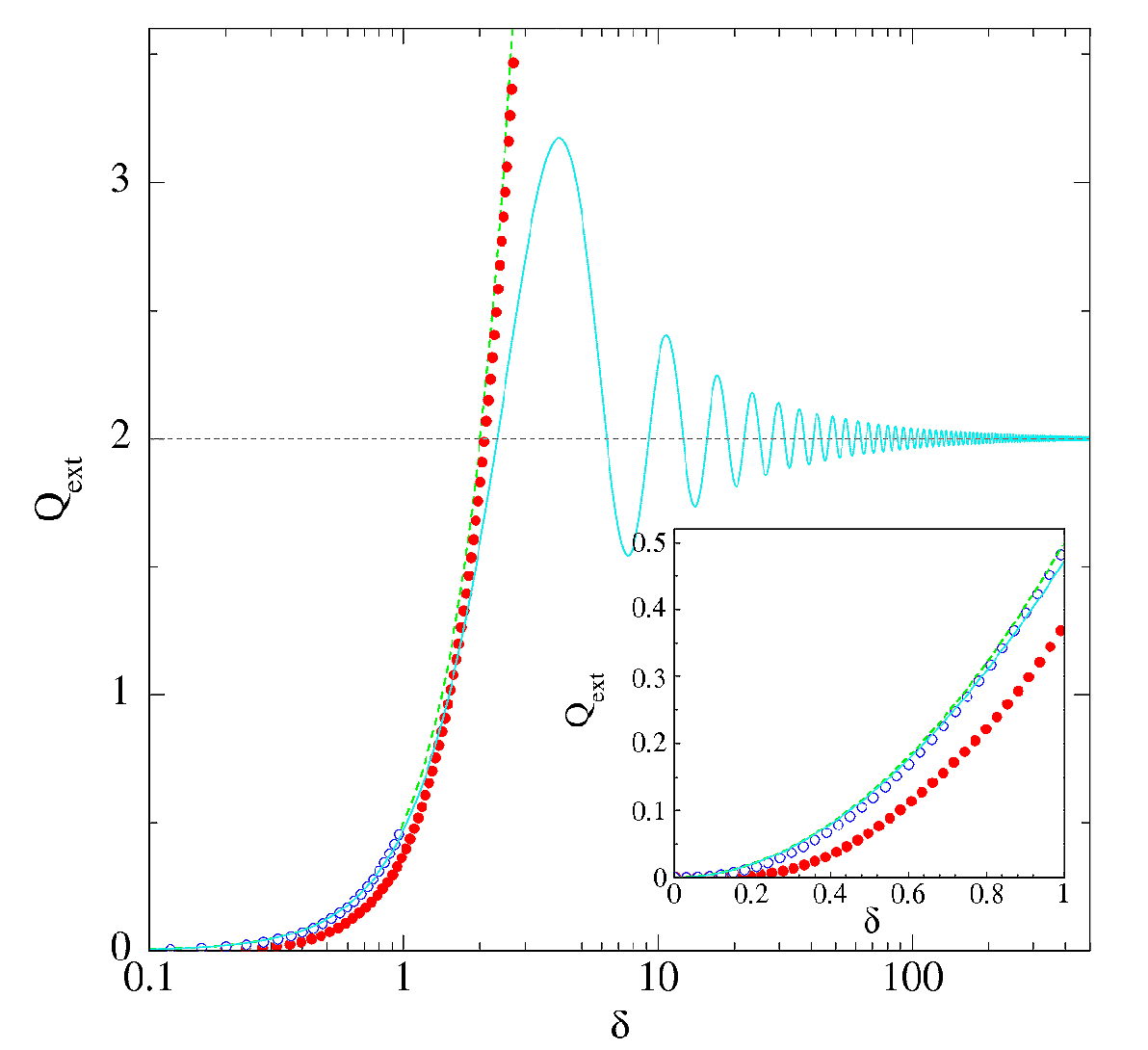}
\caption{Comparison  between of the efficiency factor $Q_{ext}^0$
obtained from Eq.~(\ref{efficiency-Anodiff}) (full line), and the
2nd order approximation in $\Delta n_p$ from
Eq.~(\ref{efficiency2}, for $\Delta n_p = 0.05$ (open dots) and
$\Delta n_p = 0.2$ (full dots). The broken line shows the
Rayleigh--Gans approximation $Q_{ext}^0 = \delta^2 /2$. The region
with $\delta \le 1$ is expanded in the inset.} \label{f1}
\end{figure}
The efficiency factor obtained from
Eq.~(\ref{efficiency-Anodiff}), whose complex oscillating behavior
can be regarded as the effect of the interference between the
transmitted and the diffracted fields, is contrasted in
Fig.~\ref{f1} with the full numerical solution of
Eq.~(\ref{efficiency2}) in the absence of inter-particle
correlations, $Q_{ext}^{\,0} =
   (2\Delta n_p/3\pi)\,C_i^{\,0}\, \delta$. The plot shows
that, for $\Delta n_p = 0.05$, the latter is very close to the RG
limit given by Eq.~(\ref{Mie_large}). As shown in the inset, the
range of validity of our 2nd order approximation shows extends up
to values of $\delta \simeq 1$  or, equivalently, for values of
the efficiency factor $Q_{ext}^{\,0}\lesssim 0.5$. Notice,
however, that for the larger value $\Delta n_p = 0.2$, included
for later convenience, differences are more marked.

Luckily, the evaluation of \emph{dispersion} effects does not
arguably suffer from this limitation. Indeed, the expression
refractive index in~(\ref{Mie_large}) does \emph{not} depend on
the particle size, and should give the correct limiting behavior
for $x\rightarrow \infty$ (at 2nd order in $\Delta n_p$). This is
also suggested from the limiting behavior of the refractive index
obtained from~(\ref{Anodiff}) using Eq.~(\ref{n_noninteracting}):
$$ n = 1+\phi \Delta n_p -(2/5)\,x^2(\Delta n_p)^3 + \mathcal{O} (\Delta n_p^5),$$
which does not contain terms in $(\Delta n_p)^2$, and depends on
particle size only at order $(\Delta n_p)^3$ and higher. Trusting
this ansatz, in what follows we shall mainly focus on the effect
of interparticle interactions on the refractive index of the
suspension, limiting the discussion of extinction properties to
dispersions of particles with a size $a \lesssim\lambda /(4\pi
\Delta n_p)$.

\subsection{Refractive index of interacting colloids: an exact limit}
\label{interacting} For small particles, including interparticle
interactions does not substantially modify the behavior of the
refractive index given by Eq.~(\ref{n_small}), since the real part
of the correlation factor is still found to be proportional to
$x^2$. Yet, $C_r$ specifically depends on the nature of
interparticle forces: the case of hard-sphere suspensions will be
discussed in Section~\ref{HS}.

Remarkably, however, in the opposite limit of $ka\to\infty$ the
real part of the correlation integral can be analytically
evaluated at \emph{any} particle volume fraction. In fact, in this
limit Eq.~(\ref{cc}) becomes
\begin{equation}
C_r= \frac{7v}{6\pi}\int_0^\infty dq\,q^2 F(q)^2 \left [ 1+
\rho_ph(q)\right ]
\end{equation}
which, by use of the convolution theorem can be written as
\begin{equation}
C_r= \frac{7\pi}{3v}\int d\br\int d\br^\prime\,
\theta(r-a)\,\theta(r^\prime-a) \left [ \delta(\br-\br^\prime) +
\rho_p \,h(\br-\br^\prime)\right ]
\end{equation}
where use has been made of the definition of the form factor
$F(q)$ in (\ref{form}). The domain limitation induced by the
presence of the $\theta$ function, implies that $|\br-\br^\prime|
< 2a$ and therefore $h(\br-\br^\prime)=-1$ in the whole
integration domain for colloids provided of a hard core
contribution. This immediately yields:
\begin{equation} \label{correffect_large}
C_r = \frac{7\pi}{3}(1-\phi) = (1-\phi)C_r^{\,0}
\end{equation}
When this result is substituted in the general expression
(\ref{ncolle_2}) for the real part of the refractive index, we
obtain the \emph{exact} limit of $n$ for $ka\to\infty$ to second
order in the particle polarizability:
\begin{equation}\label{n_large}
    n = 1+\phi \Delta n_p  + \phi(1-\phi)(\Delta n_p)^2
\end{equation}
Note that this asymptotic result is valid for \emph{any} specific
form of the interparticle interactions, provided the latter
contain a hard core contribution.

\subsection{Inclusion of the solvent}
\label{solvent} The former results have been obtained for
particles suspended in a vacuum. Nevertheless, once the refractive
index has been expressed in terms of continuum electrodynamics
quantities such as $n_p$, inclusion of the effects of a solvent,
acting as a homogeneous, non-absorbing medium of refractive index
$n_s$, is straightforward. Eq.~(\ref{n_small}) and~(\ref{n_large})
retain indeed their validity provided that we simply make the
substitutions $n\rightarrow n/n_s$, $n_p \rightarrow n_p/n_s$.
Besides, in the presence of the solvent the incident wave-vector
should be written as $k= 2\pi n_s/\lambda$. Putting $\Delta n_{ps}
= n_p -n_s$, the general expression for the complex refractive
index in Eq.~(\ref{ncolle_2}) becomes:
\begin{equation}\label{ncolle_solv}
    \frac{\tilde{n}}{n_s} = 1+\phi\,\frac{\Delta n_{ps}}{n_s}  +
    \frac{\phi}{2}\left[\frac{\phi-1}{3}+\frac{\widetilde{C}}{\pi} \right]
    \left(\frac{\Delta
    n_{ps}}{n_s}\right)^2.
\end{equation}
Note that Eq.~(\ref{ncolle_solv}) is a 2nd order expansion in
$\Delta n_{ps}/n_s$, which does not require $n_p - 1\ll1 $ and
$n_s-1\ll 1$ separately. In the limits $ka = 0$ and $ka
\rightarrow \infty$ we have therefore:
\begin{equation}\label{n_small_solv}
    \begin{array}{lll}
      n = n_s +\phi\Delta n_{ps} -\dfrac{\phi(1-\phi)}{6n_s}\,(\Delta
    n_{ps})^2  && (ka =0) \\
    \end{array}
\end{equation}
\begin{equation}\label{n_large_solv}
    \begin{array}{lll}
      n = n_s +\phi\Delta n_{ps} +\dfrac{\phi(1-\phi)}{n_s}\,(\Delta
    n_{ps})^2  && (ka \rightarrow \infty) \\
    \end{array}
\end{equation}
A note of caution is however appropriate, since the continuum
electrodynamics approach fully neglects fluctuations. It is then
worth wondering whether this simple way to account for the
presence of the solvent holds true also in the presence of
correlations, by considering again the problem in a microscopic
perspective. This is done in Appendix~\ref{solvent}, where we
explicitly show that Eq.~\ref{n_small_solv} is rigorously true
only provided that the solvent is regarded as a uniform,
uncorrelated dielectric medium.

\subsection{Effective medium approach}
\label{EffMedium}

A colloidal suspension of particles at volume fraction $\phi$ in a
solvent at volume fraction $1-\phi$ is actually a composite
medium. It is then useful to try and frame our results within the
problem of ``homogenization'' of a heterogeneous medium, which
basically consists in mapping the latter into a homogeneous
structure  by defining ``effective'', global material
properties.\cite{Markov1999} For what follows it is useful to
point out that most of the approaches has addressed the case where
these material properties are response functions to an external
field which is uniform, or slowly-varying over the length scales
that characterize the microscopic structure of the heterogeneous
medium. This is the case of the static dielectric constant, but
also of several other physical quantities such as the thermal and
low-frequency electric conductivities, or even of mechanical
quantities such as the elastic stress tensor.

In the case of a very dilute suspension of spherical particles in
a solvent, the problem is conceptually analogous to the discussion
of a system of uncorrelated point dipoles made in Section
\ref{s.dipoles}, provided that each particle is attributed a
polarizability per unit volume $\alpha_p = (\epsilon_p
-\epsilon_s)/(\epsilon_p+2\epsilon_s)$. It is therefore not
surprising that Maxwell, who first explicitly tackled this
problem,\footnote{Maxwell actually discussed the problem of the
electrical conductivity of a matrix containing spherical
inclusions. His results were later extended to the optical
properties of metal films by James Clerk Maxwell Garnett, who owes
his rather curious name to the admiration of his father, William
Garnett, for his own friend and mentor J. C. Maxwell.} obtained a
result that can be written, for the case of the effective
dielectric constant $\epsilon^*$ we are discussing
\begin{equation}\label{Max1}
    \frac{\epsilon^*}{\epsilon_s} = \frac{1+2\beta \phi}{1-\beta
    \phi},
\end{equation}
where $ \beta = (\epsilon_p
-\epsilon_s)/(\epsilon_p+2\epsilon_s)$, which is strictly related
to the CM equation.~\cite{Markov1999} As Maxwell already pointed
out, however, Eq.~(\ref{Max1}) is valid only at first order in
$\phi$, hence it should consistently be written:
\begin{equation}\label{Max2}
    \frac{\epsilon^*}{\epsilon_s} =
    1+3\frac{\epsilon_p-\epsilon_s}{\epsilon_p+2\epsilon_s}\,\phi +
    \mathrm{o}(\phi),
\end{equation}
A straightforward way to prove~(\ref{Max2}) consists in noticing
that, from the definition of the effective dielectric constant and
indicating with $\mathbf{E}_0$ an external uniform field, we must
have\cite{Landau1960}
\begin{equation*}
(\epsilon^* - \epsilon_s)\mathbf{E}_0 = \frac{1}{V}\int_V \D
\mathbf{r}\,
[\mathbf{D}(\mathbf{r})-\epsilon_s\mathbf{E}(\mathbf{r})]  =
\rho_p \int_v \D \mathbf{r}\,
(\epsilon_p-\epsilon_s)\,\mathbf{E}(\mathbf{r}),
\end{equation*}
where $\mathbf{E}(\mathbf{r})$ and $\mathbf{D}(\mathbf{r})$ are
the local, fluctuating electric  and displacement fields, $V$ is
sample volume, and the last equality is because the averaged
quantity differs from zero only within particle volume $v$. Then,
\emph{if} we assume that the field incident on particles coincides
with the external field (namely, if we neglect the additional
contributions due to the other particles), the field inside a
dielectric sphere is also uniform, and given by
$\mathbf{E}(\mathbf{r}) \equiv  [3\epsilon_s /(\epsilon_p
+2\epsilon_s)]\mathbf{E}_0$, wherefrom Eq.~(\ref{Max2})
immediately follows.

In the presence of correlations, expressions which are valid to
higher order in $\phi$ can be found only for specific geometries,
although rigorous upper and lower limits for $\epsilon^*$, such as
the Hashin-Shtrikman bonds, can be given.\cite{Markov1999} A very
interesting situation is however that of a ``weakly
inhomogeneous'' medium, which for the present purposes we identify
with a suspension of colloidal particles made of a material with
dielectric constant $\epsilon_p$, which does not differ too much
from the dielectric constant $\epsilon_s$ of the suspending
medium. Denoting by \mbox{$\overline{\epsilon} = \phi \epsilon_p +
(1-\phi) \epsilon_s$} the volume average of the dielectric
constants (which is the expression at lowest order in $\Delta
\epsilon_{ps} = \epsilon_p -\epsilon_s$ for the dielectric
constant of the mixture), and by
\mbox{$\overline{(\delta\epsilon)^2} =
\overline{\epsilon^2}-(\overline{\epsilon}\,)^2 =\phi
(1-\phi)(\Delta \epsilon_{ps})^2$} its mean square fluctuation,
one finds, at second order in $\Delta \epsilon_{ps}$
,\cite{Braun1955,Landau1960,Markov1999}
\begin{equation}
\epsilon^* = \overline{\epsilon} -
\frac{\overline{(\delta\epsilon)^2}}{3\overline{\epsilon}} =
\overline{\epsilon}  -
\frac{\phi(1-\phi)}{3\overline{\epsilon}}\,(\Delta
\epsilon_{ps})^2.\label{emix}
\end{equation}
Notice that this expression, originally derived by
Braun\cite{Braun1955} using an approach closely resembling the one
we used in Section~\ref{s.dipoles}, is valid \emph{whatever} the
spatial correlations of the particles and, in particular, for any
value of the particle volume fraction $\phi$. Remarkably,
Eq.~(\ref{emix}) also coincides with the 2nd order expansion in
$\Delta \epsilon_{ps}$ of Eq.~(\ref{Max1}), a result which has
however been derived in the \emph{uncorrelated}, single-particle
limit $\phi \rightarrow 0$. This means that, at this order of
approximation in $\Delta \epsilon_p$, the static dielectric
constant is not affected by correlations.\footnote{It is useful to
notice that, at 2nd order in $\Delta \epsilon_p$, also the
rigorous upper and lower limits for $\epsilon^*$ given by the
Hashin-Shtrikman bonds coincide.}

As we anticipated, however, Eq.~(\ref{emix}) requires the applied
electric field to be \emph{slowly-varying} on the microscopic
structural length scales of the suspension (the particle size, or
in general the correlation length for interacting particles): it
is then very useful to investigate whether Eq.~(\ref{emix}) still
holds at \emph{optical} frequencies, namely, for the refractive
index $n = \sqrt{\epsilon}$. This is readily found to be true in
the limit $ka \rightarrow 0$, where, according to Eq.~(\ref{small
ka}), $C_r^{\,0}$ vanishes as $(ka)^2$: it is indeed easy to show
that Eq.~(\ref{emix}), written in terms of the refractive indices
$n_p = \sqrt{\epsilon_p}$, $n_s = \sqrt{\epsilon_s}$, and expanded
at second order in $\Delta n_{ps}$, coincides with
Eq.~(\ref{n_small_solv}). Hence, at 2nd order in the
polarizability difference, the refractive index of a suspension of
particles small compared to the wavelength satisfies the
Lorentz-Lorenz equation at \emph{any} volume fraction. According
to our results, this does not hold true for finite values of $ka$,
where system-specific effects of the intra- and inter-particle
correlations should be expected. Remarkably, however, a distinct
limiting behavior, which is still independent from the nature and
strength of particle interactions (provided that the latter have a
hard-core contribution) and valid for any volume fraction, is
reached at large $ka$. Notice in particular that not only the
amplitude, but also the \emph{sign} of the quadratic correction in
Eq.~(\ref{n_large_solv}) differs from the CM expression.
Eq.~(\ref{n_large_solv}) is then a very general result for the
effective refractive index of a weakly inhomogeneous 2-components
medium that, at variance with Eq.~(\ref{n_small_solv}), applies
when the field varies on much shorter spatial scales than the
microscopic correlation length of the system. The fact that is
does not depend on the structural organization of the  medium, but
only on the volume fractions of the two components, suggests that
is should also be obtained from phenomenological but more general
arguments.

\subsection{Correlation effects on the refractive index for hard spheres}
\label{HS}
\begin{figure}[h!]
\includegraphics[width=\textwidth]{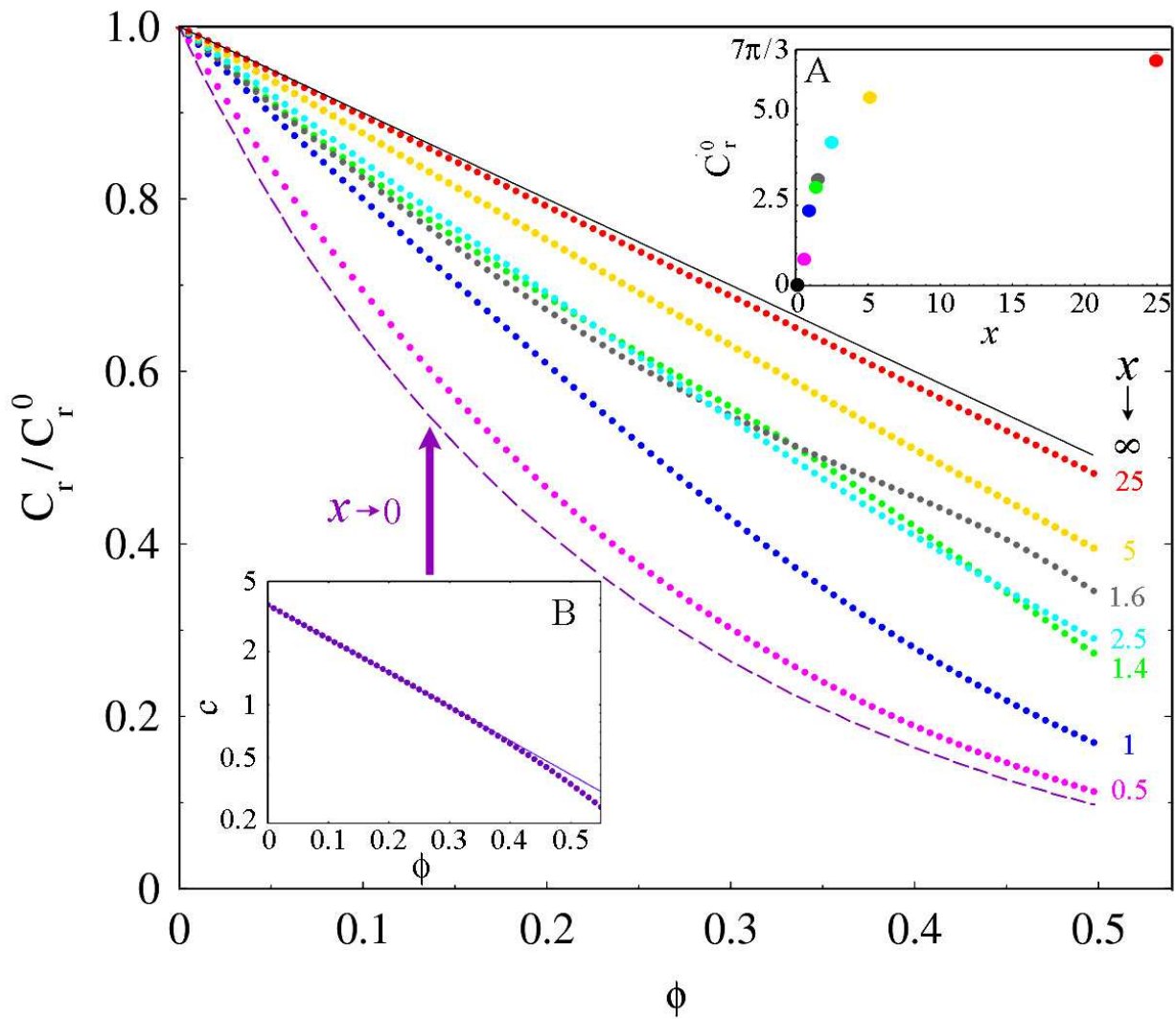}
\caption{Inset A: Single-particle (Mie) limit of the real part
$C_r^{\,0}$ of the correlation factor of hard spheres for several
values of $x =ka$. Body: Full correlation contribution $C_r$,
scaled to $C_r^{\,0}$ and plotted as a function of $\phi$ for the
same values of $x$. The full and dotted line respectively show the
limiting behavior for $x\rightarrow \infty$ and $x\rightarrow 0$.
Inset B: Slope of $C_r$ versus $x^2$ in the limit $x \to 0$,
plotted on a semi-log scale and fitted with a single exponential.}
\label{f2}
\end{figure}
For intermediate values of $x$, correlation effects on
the refractive index become system-specific: it is particularly
instructive to examine these effect for a fluid of monodisperse
hard spheres of radius $a$. Consider first the single-particle
(Mie) limit discussed in Section~\ref{Mie}, where only
intra-particle correlations are taken into account. The inset A in
Fig.~\ref{f2} shows that, in agreement with Eq.~(\ref{small ka})
and (\ref{C_p large x}), the real part $C_r^{\,0}$ of the
correlation factor, which vanishes for $x \rightarrow 0$ (the
``Clausius--Mossotti'' limit), progressively grows with $x$,
asymptotically approaching the value $C_r^{\,0} =7\pi/3$. As we
already mentioned, even in the presence of inter-particle
interactions $C_r$ retains, for small values of $ka$, a quadratic
behavior, $C_r = cx^2$. For hard spheres, Inset B shows that, to a
good degree of approximation, the slope $c$ decreases
exponentially up to $\phi \simeq 0.4$, starting from the value
$C_r^{\,0} =88\pi/75$ given by Eq.~(\ref{small ka}). The
fractional contribution $C_r /C_r^{\,0}$ due to
\emph{inter}-particle correlation is conversely shown in the body
of Fig.~\ref{f2} as a function of the particle volume fraction,
for several values of $x$. Starting from the limiting behavior
shown in inset B (dotted line), $C_r /C_r^{\,0}$ is seen to
rapidly approach, by increasing $x$, the asymptotic behavior
$C_r/C_r^{\,0} =1-\phi$ given by Eq.~(\ref{correffect_large}). For
$x \gtrsim 5$, as a matter of fact, $C_r$ is a remarkably linear
function of particle volume fraction, showing that for large $x$
only the Mie contribution and excluded volume effects are
relevant. For $1.5 \lesssim x \lesssim 2$, however, the trend of
$C_r/C_r^{\,0}$ versus $\phi$ is rather peculiar: for instance,
the curve for $x =1.6$, which at low $\phi$ lies below the curve
for $x =2.5$ as expected, crosses the latter at $\phi \simeq 0.3$,
reaching a consistently higher value at the maximum packing
fraction $\phi \simeq 0.5$ of the stable fluid phase.
\begin{figure}[h!]
\includegraphics[width=\textwidth]{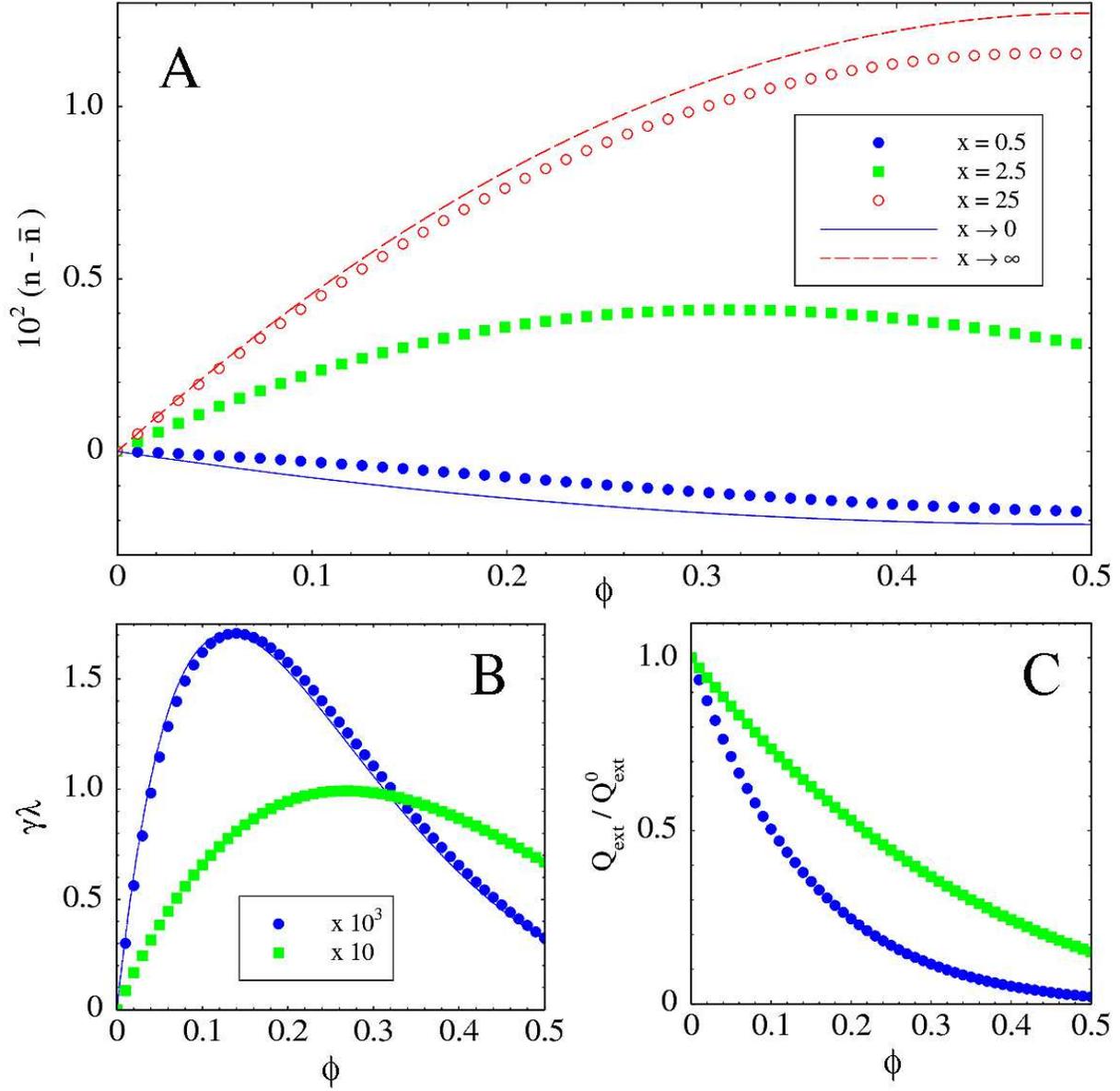}
\caption{\label{f3} Panel A: Excess correlation contribution
$n-\overline{n}$, where $\overline{n} = n_s +\Delta n_{ps} \phi$,
to the real part of the refractive index for suspensions of
polystyrene particles ($n_p =1.59$) in water ($n_s=1.33$),
corresponding to the values of $x$ in the legend. The full and
dashed lines respectively correspond to the limits $x \rightarrow
0$ in Eq.~(\ref{n_small_solv}) and $x \rightarrow \infty$ in
Eq.~(\ref{n_large_solv}). The dependence on volume fraction of the
dimensionless extinction coefficient, $\gamma \lambda$, and of the
scattering efficiency scaled to the Mie value,
$Q_{ext}/Q_{ext}^0$, are shown in Inset B and C, respectively.}
\end{figure}

These distinctive structural effects are better investigated by
considering a specific case, which also allows to inquire how
consistent and experimentally detectable are correlation
contributions to the refractive index. As a colloidal system of
practical relevance, we focus on suspensions of monodisperse
polystyrene (PS) latex particles ($n_p \simeq 1.59$) in water
($n_s \simeq 1.33$). For this colloid, despite a substantial
refractive index mismatch between particle and solvent, the terms
we neglect should not be larger than 20\% of the 2nd order term in
$\Delta n_{ps}/n_s$. Panel A in Fig.~\ref{f3} shows that, for $x =
 0.5$ ($a\simeq 0.06 \lambda$), the
difference $n- \bar n$ between the suspension refractive index and
the value obtained by simply averaging over volume fractions,
 $\bar n = n_s +\Delta n_{ps}\phi$, is pretty close to the
quadratic term in Eq.~(\ref{n_small_solv}), whereas for $x = 25$
($a\simeq 3 \lambda$) it already approaches the limiting
expression given by Eq.~(\ref{n_large_solv}). Panel A also shows
that for $x=2.5$ ($a\simeq 0.3 \lambda$), a value sufficiently
large for $C_r /C_r^0$ to show a linear trend in $\phi$ (see
Fig.~\ref{f2}), $n-\bar n$ is a quadratic function of $\phi$, as
expected from Eq.~(\ref{ncolle_solv}).

For what concerns extinction, the value $x=25$ ($\delta \simeq 10$
) is far too large to be discussed within our approximation.
Fig.~\ref{f1} shows however, that this is still reasonably
feasible for $x =2.5$ ($\delta \simeq 1$), which can then be
compared to the behavior for $x = 0.5$ ($\delta \simeq 0.2$),
corresponding to particles which are much smaller than the
wavelength. For particles of this size, the scattered intensity is
basically independent from the scattering wave-vector $q$, and
proportional to the product of the volume fraction times the
osmotic compressibility of the suspension. Using the
Carnahan-Starling equation of state for hard
spheres,\cite{Carnahan1969} both $\sigma_{ext}$ and $\gamma$
should then be proportional to
\begin{equation}\label{scatt-small HS}
    \phi \left(\frac{\partial \Pi}{\partial \phi}\right)^{-1}= \phi+\frac{2\phi^2(4-\phi)}{(1-\phi)^4}
\end{equation}
Panel B in Fig.~\ref{f3}, where this functional behavior is
compared to the dimensionless quantity $\gamma \lambda$, shows
that this is indeed the case, to a good degree of approximation.
Notice in particular that the extinction coefficient displays a
strong maximum for a particle volume fraction which is very close
to the known value $\phi \simeq 0.13$ where the scattering from
small hard-spheres peaks. A similar non-monotonic trend is
observed for $x =2.5$ too, but the value where $\gamma$ peaks is
shifted to the consistently higher value $\phi \simeq 0.27$. As we
shall shortly investigate in more detail, $\gamma$ is eventually
determined by the that part of the structure factor that is
detected within the experimentally accessible $q$-range, and this
strongly depends on particle size. The strong effect of
interparticle interactions on extinction is better appreciated in
Panel C, where we plot the volume fraction dependence of the ratio
of the scattering efficiency $Q_{ext}$ obtained form
Eq.~(\ref{efficiency2}) to its value $Q_{ext}^{\,0}$ in the
absence of inter-particle correlations. For both values of $x$,
$Q_{ext}$ strongly decreases with $\phi$, reaching a value at
$\phi =0.5$ that is about seven times smaller than $Q_{ext}^{\,0}$
for $x =2.5$, and as much as \emph{fifty} times smaller for $x =
0.5$.

\begin{figure}[h!]
\includegraphics[width=\textwidth]{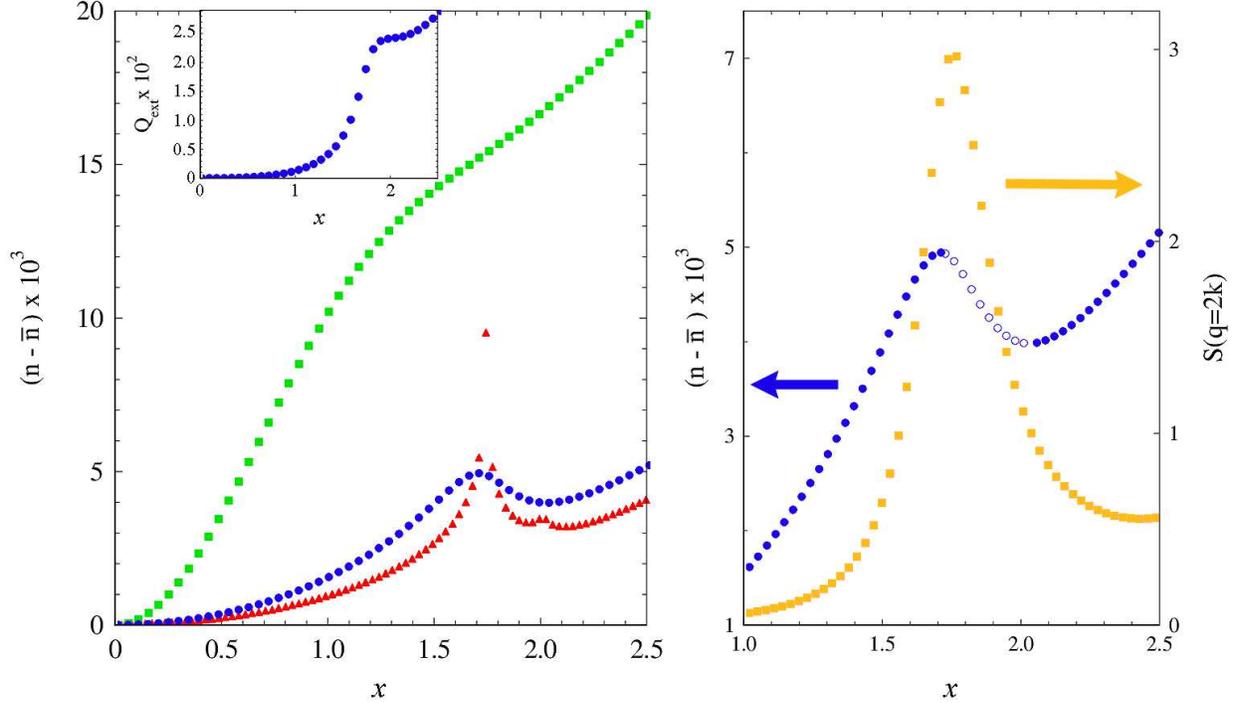}
\caption{\label{f4} Panel A: Refractive index increment $\Delta n
= n-\bar{n}$ for a suspension of polystyrene particles at $\phi =
0.50$ (fluid phase, dots) and at $\phi = 0.55$ (FCC colloidal
crystal, triangles), compared to the the values $\Delta n =
n^0-\bar{n}$ obtained by neglecting inter-particle structural
correlation (Mie limit, squares). The scattering efficiency
$Q_{ext}$ is shown in the Inset. Panel B: Detailed behavior of
$\Delta n$ in the region $ 1\le x \le 2.5$, compared to the
behavior of the structure factor of the HS fluid, calculated using
the Verlet--Weis approximation\cite{Hansen2013} and evaluated at
the maximum experimentally detectable wave-vector $q=2x/a = 4\pi
n_s/\lambda$ (see text).}
\end{figure}To highlight distinctive correlation effects, it is particularly
useful to investigate the behavior of the refractive index for
$\phi = 0.5$, which is the limiting volume fraction of a hard
spheres fluid, as a function of the scaled particle size $x$. In
Panel A of Fig.~\ref{f4}, the difference $\Delta n$ between the
refractive index $n$ and its volume-average approximation $\bar{n}
= n_s + \Delta n_{ps}$ is contrasted to the single-particle Mie
limit $\Delta n = n^0- \bar{n}$. For comparison, we also include
the corresponding data for the face-centered-cubic crystal at
$\phi = 0.55$ in equilibrium with the fluid phase. It is important
to point out that the latter assumes orientation symmetry, and
therefore corresponds to the data that would be obtained for a
randomly-oriented polycrystalline sample. Several features of the
plot are worth be pointed out. First, $n-\bar{n}$ is
\emph{substantially} smaller for both the fluid and colloidal
crystal phase than in the non-interacting approximation. In
particular, the quadratic increase of $\Delta n_p$ at small $x$,
expected from the discussion of $C_r$ in Fig. \ref{f2}, is  more
than one order of magnitude weaker than in the Mie case. The
effect of repulsive HS interactions is even more dramatic on
extinction: the Inset in Panel B shows indeed that, in the fluid
phase (extinction in the crystal vanished, since it is assumed as
ideal), the scattering efficiency is extremely small for $x
\lesssim 1$, rising for $x\gtrsim 2$ to values which are still
several times smaller than in the Mie theory. A very peculiar
feature is finally the presence of  a strong peak for the crystal
phase, and of a less pronounced ``bump'' for the fluid phase,
occurring in the region $1.5 \lesssim x \lesssim 2$ where we
already pointed out a peculiar behavior of $C_r$ (see
Fig.~\ref{f2}). The origin of this rather surprising effect can be
grasped by considering Panel B . There, together with an expanded
view of the ``bump'' region, we plot the structure factor $S(q)$
for a HS fluid at $\phi =0.50$, calculated at the wave-vector
$q=2k = 4\pi n_s/\lambda$ corresponding to the
\emph{backscattering} condition $\theta = \pi$.  In other words,
for a given experimental value $x^*$, fixed by both the particle
size and the incident wavelength, only those wave-vectors $q$ of
the structure factor with $q \le 2x^*/a$ fall within the
detectable range and contribute to the scattering cross section.
>From the plot, we clearly see that the refractive index increase
is associated with (and actually slightly anticipates) the
progressive ``entrance'' of the first peak of $S(q)$ in the
detectable range. A further interesting observation comes from
noticing that, for a fixed particle size, the curve is basically a
plot of the refractive index versus the inverse of the incident
wavelength. Then, the trailing part of the peak which follows the
maximum, shown with open dots in Panel B, corresponds to a region
where the refractive index increases with increasing $\lambda$,
which is the hallmark of an \emph{anomalous dispersion}
region.\footnote{From Panel A, anomalous dispersion is of course
present, and even much more pronounced, for the crystal phase.
However, since we simply model the system as an FCC crystal at
$T=0$, neglecting therefore the ``rattling'' motion of the
particles at finite $T$, we cannot make a realistic correlation
with the behavior of $S(q)$. Notice however that, in spite of the
fact that in this idealized model the structure factor peaks are
discontinuous delta functions, the peak in the refractive index is
finite.} As a matter of fact, the overall trend strongly resembles
the behavior of the refractive index $n(\omega)$ of a Lorentz
oscillator close to its natural resonance frequency $\omega_0$:
$n$ already shows an increase for $\omega < \omega_0$, followed by
an anomalous dispersion region where $n(\omega)$ is a decreasing
function of $\omega$, and by a final recovery. No true absorption
is however present in the problem we are considering. This finding
seems to suggest that, besides in resonant absorption, anomalous
dispersion may take place in the presence of any process in the
medium, such as scattering, that lead to extinction of the
incident field.

\subsection{Feasibility of the experimental determination of correlation effects}
\label{exper}It is useful to inquire whether and how correlation
effects on the refractive index can be experimentally
investigated. In Panel A of Fig.~\ref{f3} the correlation
contribution to $n$ reaches, for $x=2.5$ and $\phi \gtrsim 0.3$, a
value of about $4 \times 10^{-3}$, which is well within the
accuracy of a good refractometer. However, Panel B shows that in
these conditions extinction is quite large, giving an extinction
length $\gamma^{-1} \simeq 10 \lambda \simeq 30a$: investigations
should then be performed using method exploiting a low penetration
depth (see below). Given the low extinction, correlation effects
are arguably much easier to be detected in the very large $\phi$
limit discussed in Fig.~\ref{f4}, . For instance, for $x= 1$,
where in fluid phase $Q_{ext} \simeq 10^{-3}$, corresponding to an
extinction length still as large as about $300\lambda$, the
refractive indexes of both the fluid and the crystal phase already
differ from the Mie prediction by about $10^{-2}$, and of $6\times
10^{-4}$ between themselves.

Unfortunately, accurate data on the refractive index of dense
colloids are scarse, and not very recent.\cite{Meeten1991,
Mohammadi1995} Moreover, as common in light scattering practice,
experiments mostly focused on measuring the refractive index
increment $\D n/\D \phi$ (or, more usually $\D n/\D c$, where $c$
is the concentration in mass/volume), which amounts to implicitly
assume that $n$ is linear in $\phi$.  However, at first order in
$\phi$, the correlation factor $C_r$ reduces to the Mie limit
$C_r^{\,0}$, and inter-particle correlations show up only at order
$\phi^2$. It is however worth mentioning a rather surprising
result obtained long ago by Okubo,\cite{Okubo1990} which may be
related to the results we discussed in Fig.~\ref{f4}. By measuring
the refractive index of strongly deionized charged PS suspensions,
Okubo observed indeed a substantial peak in the refractive index,
located  very close the transition between the fluid  and the
colloidal crystal phase. Unfortunately, this early investigation
has not been further pursued, at least to our knowledge.

Modern approaches based on fiber optic sensing\cite{Banerjee2007},
or made in a total internal reflection
configuration\cite{Sarov2008}, should provide a sufficient
accuracy to detect correlation effects even for strongly turbid
samples. Yet, when using methods relying on so small penetration
depths, care should be taken to avoid probing correlation effects
on the particle distribution at the interface between the solution
and the sensor wall, rather than bulk structural properties. An
interesting alternative would be using a novel optical correlation
method recently introduced by Potenza \emph{et
al.},\cite{Potenza2010} which consists in measuring the
2-dimensional power spectrum $P(q_x, q_y)$ of the transmitted beam
intensity distribution on a plane placed at close distance $z$
from the sample. By means of the optical theorem, one finds for
$N$ identical scatterers:
\begin{equation}\label{Talbot}
    P(q_x,q_y) =
    \frac{4\pi^2}{k^2}|Ns(0)|^2\sin^2\left(\frac{q^2z}{2k} -\varphi\right)
\end{equation}
where $q^2 = q_x^2+q_y^2$ and (in our notation) $\varphi =
\arg[s(0)]$. Eq.~(\ref{Talbot}) describes a fringe-like pattern
characterized by a phase shift $\varphi$ which is directly related
to the ratio between the imaginary and the real part of the
scattering amplitude. One of the major advantages of the method is
that it can be applied to very turbid samples too, since multiple
scattering yields only a constant background that can be easily
subtracted out. The technique has successfully been applied to
investigate dilute colloidal suspension.\cite{Potenza2010} At
sufficiently high $\phi$, however, inter-particle correlation
effects should yield noticeable deviations with respect to the Mie
expression used  to evaluate the fringe pattern.

\section{Conclusions}
\label{conc} In this article we have presented a general
microscopic theory for the attenuation and the phase delay
suffered by an optical plane wave that crosses a system of
interacting colloidal particles, deriving an expression for the
forward scattered wave, exact at second order in the molecular
polarizability, which explicitly takes into account the
interactions among all induced dipoles.  Whereas previously
available treatments have separately discussed either attenuation
(neglecting corrections due to radiation reaction) or refractive
index  (using some variant of  the Lorentz-Lorenz formula and
ignoring interparticle correlations), our approach treats on an
equal basis the real and the imaginary part of the refractive
index. In detail:
\begin{itemize}
    \item We have investigated the role of radiation reaction on light
extinction, showing that the structural features of the suspension
are encoded into the forward scattered field by multiple
scattering effects, whose contribution is essential for the
so-called "optical theorem" to hold  in the presence of
interparticle interactions. The local field acting on a specific
dipole is the sum of the external field plus all the fields due to
the presence of all the other oscillating dipoles within the
scattering volume. Our treatment considers the average local
field, which is polarized as the external field, while the
fluctuations of the local field, not discussed here, give rise to
what is usually called multiple scattering;
\item In the case of negligible interparticle interactions,
our results are found to be consistent, at second order in the
polarizability, with the exact Mie theory for spherical particles;
\item We have discussed our results in the framework of effective medium theories, presenting a general
result for the effective refractive index valid, whatever the
structural properties of the suspension, in the limit of a
particle size much larger than the wavelength;
\item In the case of correlated particles we found that significant corrections to the value of the
refractive index exist when the $x$ parameter is of the order of
one, that is, when the particle size is comparable to the
wavelength of light;
\item By treating concentrated hard-sphere suspensions, we have unraveled subtle
anomalous dispersion effects for the suspension refractive index
and we have discussed the feasibility of an experimental test of
our calculations.
\end{itemize}

It is finally useful to point out that the general approach we
have followed can in principle be extended to investigate other
interesting physical problems. Strong analogies exist for instance
between the scattering of (vector) electromagnetic and (scalar)
ultrasonic waves from a particle dispersion. Although in the case
of ultrasonic scattering no analogous of point-like dipoles
exists, an approach formally identical to the Mie scattering
theory can be developed, once the whole colloidal particle is
assumed to be an elementary scatterer, responding to the incident
acoustic field via its density and compressibility difference with
the solvent.\cite{Morse1968} An extension of the approach we
developed for the refractive index might then provide an explicit
expression for the dispersion of the sound speed in a correlated
suspension. It is however worth pointing out that, in acoustic
scattering, absorption effect are usually far from being
negligible.

Similarly, finding the thermal conductivity of a suspension within
an effective medium approach is formally analogous to evaluate its
dielectric constant in the long-wavelength limit. In
Section~\ref{EffMedium} we have shown that, for a weakly
inhomogeneous medium, the static dielectric constant is not
affected by correlation, and is always given by Eq.~(\ref{emix}):
the same result should then hold for the thermal conductivity.
However, the case of a dispersion of correlated particles with a
thermal conductivity much higher than the base fluid (metal
nanoparticles, for instance) could still be investigated by a
suitable extension of the general equations (\ref{eqn},
\ref{intc}), at least numerically. This may shed light on the
highly debated problem of the so-called ``anomalous'' enhancement
of thermal conductivity in nanofluids.~\cite{Eapen2010}
\appendix
\section{}
 \label{append}
For a homogeneous, uncorrelated mixture of point-like particles
with polarizabilities $\alpha_1$, $\alpha_2$ and number densities
$\rho_1$, $\rho_2$, a straightforward generalization of
Eq.~(\ref{eq3}), applied to a slab geometry for an incident field
of the form (\ref{ext}), yields, at 2nd order in the
polarizability:
\begin{equation}
n^{d} = 1 + 2\pi(\alpha_1\rho_1+ \alpha_2\rho_2) +\frac{2}{3}\pi^2
(\alpha_1\rho_1+ \alpha_2\rho_2)^2, \label{cm3}
\end{equation}
where the superscript ``$d$'' is to remind that Eq.~(\ref{cm3}),
which just states that in the absence of correlation the
polarizability per unit volume of the mixture is additive, applies
only to  point-like dipoles.  Let us now identify species $2$ with
molecules constituting the colloidal particle. When correlations
are included, by generalizing the procedure discussed in Section
\ref{colloid}, we obtain a general expression of the complex
refractive index of colloidal particles embedded in a correlated
fluid:
\begin{eqnarray}
\tilde n = 1 &+& 2\pi(\alpha_1\,\rho_1+ \alpha_p\,\phi)
+\frac{2}{3}\pi^2
(\alpha_1\,\rho_1+ \alpha_p\,\phi)^2 \nonumber \\
&+& 2\pi\left \{ \alpha_1^2\, \rho_1^2\, C[h_{11}(q)] +
\alpha_p^2\,\phi v C[F^2(q) S_{pp}(q)] + 2\alpha_1 \alpha_p \,
\rho_1\phi\,C[ F(q)h_{1p}(q)] \right \} \label{mix}
\end{eqnarray}
where we have defined a correlation functional $C[f(q)]$ that
generalizes expression (\ref{intc}) to:\begin{equation} C[f(q)] =
\frac{1}{4\pi^2 k^2}\,\int d\bq \, \left [
\frac{k^4+(\bk\cdot\bq)^2}{q^2 - k^2 +i\eta} - \frac{k^2}{3}\right
] \, \,f(|\bk-\bq|)
\end{equation}
In the special limit where we identify type-$1$ particles with the
molecules of the solvent, considered as an incompressible
continuum, the number density of the solvent $\rho_s$ in the
\emph{free} volume $V(1-\phi)$ is related to $\rho_1$ by
\begin{equation}
\rho_s = \frac{\rho_1}{1-\phi}
\end{equation}
and the correlation functions $h_{11}(q)$, and $h_{1p}(q)$ can be
related to that of the colloidal particle $h_{pp}(q)\equiv h(q)$:
\begin{eqnarray}
h_{11}(q) &=& \frac{\rho_p v^2}{(1-\phi)^2} \,F^2(q) [1+\rho_p\,h(q)] \\
h_{1p}(q) &=& -\frac{v}{1-\phi} \,F(q) [1+\rho_p\,h(q)]
\end{eqnarray}
When these expressions are inserted into Eq. (\ref{mix}) we find
\begin{equation}
n = 1 + 2\pi\left [\alpha_s\rho_s(1-\phi)+ \alpha_p \,\phi \right
] + \frac{2}{3}\,\pi^2 \left [\alpha_s\rho_s(1-\phi)+ \alpha_p
\,\phi \right ]^2. \label{nmix}
\end{equation}
It is easy to show that this form is in fact fully equivalent to
Eq.~(\ref{ncolle_solv}), when the latter is expanded to second
order in $n_s-1$ and $n_p-1$.

\end{document}